\documentclass[twocolumn,showpacs,preprintnumbers,amsmath,amssymb]{revtex4-1}

\pdfoutput=1 
\usepackage{graphicx}
\usepackage{dcolumn}
\usepackage{bm}
\usepackage{multirow}
\usepackage[caption=false]{subfig}

\usepackage{float}
\usepackage{enumitem}
\graphicspath{{./detector_plots/}{./eA_plots/}{./ep_plots/}{./HERA_comp/}{./misc_images/}}
\begin{document}


\title{Exclusive vector meson production at an electron-ion collider}
\author{Michael Lomnitz and Spencer Klein} \affiliation{Lawrence Berkeley National Laboratory, Berkeley CA, 94720, USA}
 
\date{\today}

\begin{abstract}

Coherent exclusive vector meson electroproduction is a key physics channel at an electron-ion collider.  It probes the gluon structure of nuclei over a wide range of $Q^2$, and can be used to measure nuclear shadowing and to search for gluon saturation and/or the colored glass condensate.   In this paper, we present calculations of the kinematic distributions for a variety of exclusive vector meson final states: the $\rho$, $\phi$, J/$\psi$, $\psi'$ and the $\Upsilon$ states.  The cross-sections for light and $c\overline c$ mesons are large, while $\Upsilon$ states should be produced in moderate numbers at a medium energy  EIC (the proposed U.S. designs) and in large numbers at the LHeC.    We investigate the acceptances for these states, as a function of detector rapidity coverage.  A large-acceptance detector is needed to cover the full range photon-nucleon collision energies produced at an EIC; a forward detector is required to observe vector mesons from the most energetic photon interactions, and thereby probe gluons at the lowest possible Bjorken-$x$ values.  
\end{abstract}
\maketitle

\section{Introduction}

Coherent vector meson (VM) photoproduction and electroproduction are important channels for investigating the composition of protons and nuclear targets \cite{Accardi:2012qut}.  The reactions have large cross-sections and simple final states, and are expected to play a key role at an upcoming Electron-Ion Collider (EIC), where they will be used to study the parton structure of protons and nuclear targets.   

In coherent VM photoproduction an incident photon fluctuates to a quark-antiquark ($q\overline q$) pair which then scatters elastically from the target, emerging as a VM.  The scattering occurs via the exchange of a color-neutral object, either a meson (at low photon energies) or the Pomeron (for higher photon energies).   In some calculations, the meson exchange is treated as a Regge-pole exchange \cite{Fazio:2011ex}.  In dipole models, the incident $q\overline q$ is treated as a dipole which interacts with the target \cite{Martin:1996bp,GolecBiernat:1998js,Kowalski:2006hc}.  

Because the Pomeron is mostly gluonic,  exclusive VM production cross-sections can be related to the gluon structure functions \cite{Jones:2013pga,Jones:2016ldq} using perturbative quantum chromodynamics.  The higher the photon energy, the lower the Bjorken-$x$ of the target gluon that can be probed. However, because the Pomeron consists of at least two gluons, and has internal structure, the relationship between gluon structure functions and photoproduction/electroproduction cross-sections is more complex than for reactions like photoproduction of open charm or dijets.  Analyses need to account for the second gluon in determining gluon structure functions.  Alternately, one can perform calculations in a colored-glass condensate framework \cite{Mantysaari:2017slo}.   

Most of these models produce predictions which are in good agreement with existing data from the HERA Collider, which studied the photoproduction of light and heavy (charmed and bottom) quarkonium in $ep$ collisions over a broad range of photon energies and $Q^2$.   

Unfortunately, HERA never accelerated ion targets.   Ion targets exhibit nuclear shadowing, a modification of the gluon density in nuclear targets compared to those in single nucleons.  The higher density may also lead to new phenomena, such as saturation, whereby multiple low Bjorken-$x$ gluons recombine into a single gluon.  This can happen in individual nucleons, but is more likely in heavy nuclei, where the gluon density is larger.     These phenomena can be accurately studied experimentally at an electron-ion collider by comparing the cross-sections for VM photoproduction in $ep$ and $eA$ collisions.  Many of the theoretical uncertainties cancel out in the ratio, allowing nuclear effects to be studied.  The CEBAF accelerator studies electro- and photo- production, but it is limited to a maximum photon energy of 12 GeV, sufficient only to study production light quark quarkonium on gluons with Bjorken-$x  \gtrapprox 0.1$. 

Coherent VM photoproduction on heavy ion targets has been studied with ultra-peripheral collisions, but there the $Q^2$ is  fixed by the VM mass.  $Q^2 \approx (M_V/2)^2$.  Data from STAR  \cite{Adler:2002sc, Abelev:2007nb,Adamczyk:2017vfu} and PHENIX \cite{Afanasiev:2009hy} at RHIC and from ALICE \cite{Abelev:2012ba,Adam:2015gsa} and CMS \cite{Khachatryan:2016qhq} at the LHC show that VM photoproduction is moderately suppressed compared to Glauber model extrapolations of a proton baseline \cite{Klein:2017vua}.  However, exploring how these nuclear effects scale with $Q^2$ requires an electron-ion collider.

Planned EIC studies go far beyond measurements of the cross-sections \cite{Accardi:2012qut}.   The distribution of momentum transfers from the nuclear target, $d\sigma/dt$ is sensitive to the distribution of interaction sites within the nucleus, and so is sensitive to phenomena like saturation or the colored glass condensate \cite{Toll:2012mb}.   $d\sigma/dt$ exhibits a diffractive pattern, with maxima and minima that are characteristic of the density profile of the nuclear target.   Gluon shadowing is largest in the interior of a nucleus, and smaller on the outer surfaces \cite{Kumano:1989eh,Kitagaki:1988wc,Emelyanov:1998yul}, so it should alter the distribution of interaction points within the nucleus, and so alter the shape of $d\sigma/dt$.  The transverse momentum is conjugate to transverse position, so one can use $d\sigma/dt$ into determine, via a Fourier-Bessel (Hanckel) transform the effective shape of the nuclear target \cite{Adamczyk:2017vfu,Toll:2012mb}.  

Observing these diffraction patterns in $d\sigma/dt$ in detail requires very large numbers of events; the STAR analysis \cite{Adamczyk:2017vfu} used about 300,000 events.  Tracking how $d\sigma/dt$ varies with $Q^2$ and other kinematic variables require orders of magnitude more events.  This requires both the very high luminosities planned for an EIC, and also large-acceptance detectors.

In this paper, we calculate the rates and kinematic distributions for photo-production (photon virtuality $Q^2\approx 0$) and electroproduction ($Q^2 > \approx 1$ GeV$^2$) of a variety of VMs.  We consider three possible accelerators: eRHIC proposed at Brookhaven National Laboratory \cite{eRHIC}, JLEIC proposed at Jefferson Lab \cite{JLEIC} and  LHeC proposed at CERN \cite{AbelleiraFernandez:2012cc} .  The plans for these accelerators are still evolving; we use the specifications from recent references, summarized  in Tab. \ref{tab:accel}.  We also consider collisions at the now-decommissioned HERA accelerator at DESY as a check of the calculations.  

The main goal of this paper is to present experimentally useful cross-sections and kinematic distributions at these accelerators, to facilitate designing detectors and planning an experimental program.  We consider a wide variety of VMs, including the $\rho$, $\omega$, $\rho'$, and $J/\psi$, $\psi(2S)$ and $\Upsilon$.   Our calculations take, as input, data from HERA.  However, high-energy electroproduction (at large $Q^2$) on ion targets has not been studied; in this region, we rely on theoretically informed extrapolations.

One key variable for experimental design is the pseudorapidity of the final state particles, including the daughters from the VM decay.  The decay angles account for the photon polarization.     To facilitate studies of detector acceptances, we have put our work in the form of an event generator which generates lookup tables which can then be used to quickly generate VMs and decay them to produce experimentally visible final states.   We used the framework of the STARlight Monte Carlo generator \cite{Klein:2016yzr} as a starting point.  It simulates a wide variety of VM final states, produced on a wide range of ions.  It  has been shown to do an excellent job of reproducing the cross-sections and kinematic distributions for $\rho^0$ photoproduction at RHIC \cite{Abelev:2007nb} and the LHC \cite{Adam:2015gsa}.   It underestimates the cross-sections for heavier VMs like the $J/\psi$, because it does not include gluon shadowing, but it largely reproduces the kinematics of the process \cite{Khachatryan:2016qhq,PhysRevLett.113.232504}.  We made extensive additions to the code, to accommodate photons emission over a wide photon range, to track the outgoing electron, and to correctly model the photonuclear cross-sections and decay angular distributions for a wide range of $Q^2$.  The code, dubbed eSTARlight, used here is available on request.  

\begin{table}
\begin{tabular}{|l|c|r|r|}
\hline
Accelerator & Collision & Electron & Heavy Ion       \\
                    & System    & Energy   & Energy  \\
\hline
eRHIC \cite{eRHIC}         & $ep$      &  18 GeV  & 275 GeV \\
-                                       & $eA$      &  18 GeV  & 100 GeV/nucleon \\
\hline           
JLEIC  \cite{JLEIC}         &$ep$       &  10 GeV            &   100 GeV \\ 
-                   &$eA$      &   10 GeV  &   40 GeV/nucleon  \\
\hline
LHeC     \cite{AbelleiraFernandez:2012cc}         &$ep$       &  60 GeV            &   7 TeV \\ 
-                   &$eA$      &   60 GeV            &    2.8 TeV/nucleon  \\
\hline
HERA          &$ep$       &   27.5 GeV          &   920 GeV  \\  
\hline
\end{tabular}
\caption{The characteristics for the accelerators considered here. The ion energies are per-nucleon. JLEIC and LHeC studies plan on lead beams, while eRHIC is focused on gold for its heavy ion running. 
\label{tab:accel}
}
\end{table}

This code covers similar reactions as  the Sar{\it t}re generator \cite{Toll:2012mb,Toll:2013gda}.  Sar{\it t}re has a more detailed model for coherent and incoherent production, but eSTARlight handles a wider range of final states and a more detailed model of the final state angular distributions. 

\section{Cross-section Calculations}

The cross-section for production of a VM $V$ on a target $A$ may be written as
\begin{equation}
\sigma (eA\rightarrow eAV) \int dW \int dk \int dQ^2 \frac{d^2N_\gamma}{dkdQ^2} \sigma_{\gamma^* A\rightarrow VA} (W,Q^2)
\end{equation}
where $d^2N_\gamma/dkdQ^2$ is the photon flux, $W$ is the $\gamma p$ center of mass energy, $k$ is the photon energy, $Q^2$ is the photon virtuality, $\sigma_{\gamma^* A\rightarrow VA} (W,Q^2)$ is the $\gamma A$ cross-section.    

\begin{figure}
\includegraphics[width = 0.3\textwidth]{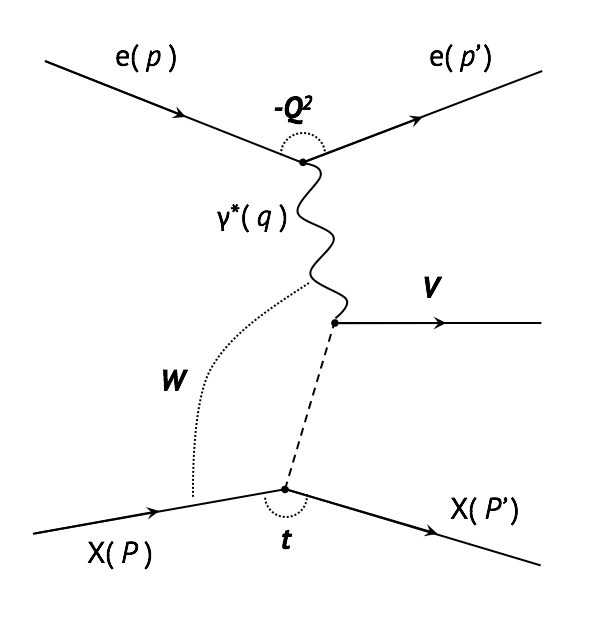}
\caption{Schematic diagram of elastic VM electroproduction $\sigma(eX\rightarrow eVX)$, illustrating the relevant kinematic variables.}\label{fig:kinematics}
\end{figure}

Figure \ref{fig:kinematics} shows a schematic diagram of elastic VM electroproduction reactions $e(p)X(P)\rightarrow e(p')V(v)X(P')$, together with some of the relevant kinematic variables:

\begin{itemize}
\item $Q^2 = -q^2 = -(p-p')^2$, the negative of the invariant mass of the exchanged photon;
\item $y = (q\cdot P)/(p\cdot P)$, the fraction of the electron energy transferred to hadronic final state in the rest frame of the initial state proton (target frame);
\item $W^2 = (q+P)^2$, the squared invariant mass of the photon-proton system;
\item $t=(P-P')^2$, the four-momentum transfer squared at the proton vertex;
\item Bjorken-$x = Q^2/(2P\cdot q)$, the momentum of the struck gluon in the target.
\end{itemize}

When $k^2\gg Q^2$ (most of the phase space), one can determine the Bjorken-$x$ values from $x=Q^2/2km_p$ \cite{Crittenden:1997yz}, where $m_p$ is the proton mass.  The most energetic photons probe the lowest $x$ values, producing VMs at far forward rapidities. 

\subsection{Photon flux}

In the frame of reference where the target is at rest, the photon flux is \cite{Budnev1975181}:
\begin{equation}
\frac{d^2N_\gamma}{dkdQ^2} = \frac{\alpha}{\pi} \frac{dk}{k} \frac{dQ^2}{Q^2} \left[1-\!\frac{k}{Ee} + \!\frac{k^2}{2E_e^2}
- \left(1- \frac{k}{Ee}\right) \!\left|\frac{Q^2_{\rm min}}{Q^2}\right|\right]
\label{eq:photon_flux}
\end{equation}
where $E_e$ is the electron energy.    The photon flux is predominantly at small values of $Q^2$,  with the minimum $Q^2$
\begin{equation}
Q^2_{\rm min} = \frac{m_e^2 k^2}{E_e(E_e-k)}
\end{equation}
which depends on the final state, while the maximum $Q^2$ is determined by the electron energy loss,
\begin{equation}
Q^2_{\rm max} = 4E_e(E_e-k).
\end{equation}
To avoid a slightly problematic endpoint region, we require $ k < E_e-10m_e$, with negligible overall effect.  

The value $Q_{max}$ is the maximum momentum exchanged allowed for a photon emitted by an electron.  In order for the photon to interact coherently with the target, the fluctuation into a $q\overline q$ dipole must live long enough to probe the entire target  \cite{PhysRevC.65.035201}. This sets a maximum value for the virtuality in terms of the coherence time (or length) of the boosted dipole:
\begin{equation}
l_c = \frac{2 k}{Q^2 + M_V^2}.
\label{eq:lc}
\end{equation}
When $l_c$ is less than the nuclear radius, $R_A$, then full longitudinal coherence is lost.  As will be discussed below, we impose a minimum requirement on $k$ to require $l_c >R_A$.
 
Unlike in the case of photo-production, exchanged virtual photons are polarized with important consequences to the angular distribution of both the final state meson and the direction of the decay products. Although many theoretical approaches are given in the literature, they are generally complex and the full treatment is outside the scope of this generator. We instead use parameterizations of existing data, as is discussed in section \ref{sec:AngularDistribution}. 

\subsection{Photonuclear cross-sections}\label{sec:photonuclear_x_sec}

The coherent photonuclear cross-sections are based on parameterizations of $\sigma(\gamma p\rightarrow Vp)$ from HERA data, as given in Ref.  \cite{Klein:1999qj}, and updated in Ref.  \cite{Klein:2016yzr}.  These cross-sections are for real photons,  We included the $Q^2$ dependence following Eq. 49 of Ref. \cite{Adloffetal.2000}:
\begin{equation}
\sigma_{\gamma^* A\rightarrow VA} (W,Q^2) = f(M_V)\sigma(W,Q^2=0) \left(\frac{M_V^2}{M_V^2+Q^2}\right)^n
\label{eq:sigmaQ2}
\end{equation}
where $\sigma(W,Q^2=0)$ is the cross-section for VM photoproduction with real photons and 
\begin{equation}
n=c_1 + c_2(Q^2+M_V^2),
\label{eq:n}
\end{equation}
where $c_1$ and $c_2$ for $\rho$ and $\phi$ are taken from fits in \cite{Aaron2010}. For $J/\psi$ the constants are determined by us by fitting the HERA data with the same functional form.  The values are given in Table \ref{tab:n}.  Typically, $f(M_V)$ is a Breit-Wigner function  \cite{Klein:1999qj}, but it may also include additional components, such as continuum direct $\pi^+\pi^-$ production.   

For some mesons, experimental data on the $Q^2$ evolution of the cross-section is lacking, so we use the values for similar mesons.  We use the same $c_1$ and $c_2$ for the $\rho'$ and $\omega$ as for the $\rho$, and take the $\psi(2S)$ and $\Upsilon$ values from the $J/\psi$, even though there are theoretical indications that the cross-sections to produce radially excited mesons have a different $Q^2$ evolution than the $1S$ levels \cite{Nemchik:1996cw,Martin:1997sh}.

For proton targets, $ \sigma(W,Q^2=0)$ may be parameterized as the sum of two power laws \cite{Breitwegetal.1999}:
\begin{equation}
\sigma(W,Q^2=0) = \sigma_P W^\epsilon + \sigma_M W^\eta.
\end{equation}
The first term is for production by Pomeron exchange, while the second is for meson exchange.  The parameters $\sigma_P$, $\epsilon$, $\sigma_M$ and $\eta$ depend on the produced meson.  For light mesons $\epsilon=0.22$ \cite{Klein:1999qj}, so the cross-section rises slowly with increasing $W$.  For heavy (charmed/bottom) mesons, $\epsilon$ is larger, and a more sophisticated parameterization is used to account for threshold effects \cite{Klein:2016yzr}.  The second, meson-exchange term is only present for the $\rho$ and $\omega$.  $\eta$ is negative, usually in the range $-1.2$ to $-1.9$, so this contribution is most important for lower photon energies.  Nevertheless, it is a significant contributor to the cross-section at medium-energy EICs.  

For heavier targets, we calculate the cross-section using a quantum Glauber calculation \cite{Frankfurt:2002wc} which accounts for the possibility of a single incident $q\overline q$ dipole interacting multiple times as it traverses the target.  This is most important for the largest dipoles, {\it i. e.} for lighter mesons with $Q^2\approx 0$.   The procedure follows Ref. \cite{Klein:1999qj}, except that its Eq. (12) is replaced with:
\begin{equation}
\sigma_{tot.}(VA) = \int d^2\vec{r}\ 2\big[1-e^{-0.5*\sigma_{\rm tot}(Vp)T_{AA}(\vec{r})}\big].
\label{eq:quantumG}
\end{equation}
where $\sigma_{tot.}(VA)$ is the total vector-meson nucleon cross-section, $\sigma_{\rm tot}(Vp)$ is the total VM-proton cross-section, and $T_{AA}(\vec{r})$ is the nuclear thickness function calculated for a Woods-Saxon nuclear density distribution. 

Compared to a classical Glauber calculation (Eq. 12 of Ref.  \cite{Klein:1999qj}) which gives a much better match to the data for $\rho^0$ photoproduction in UPCs, Eq. \ref{eq:quantumG} increases the cross-section for $\rho^0$ photoproduction by more than a factor of two; the difference is much smaller for other mesons. The difference drops as $Q^2$ rises.   The discrepancy between the quantum Glauber calculation and the data is also explainable as due to nuclear shadowing \cite{Frankfurt:2015cwa}.

At large $Q^2$, the longitudinal coherence length, Eq. \ref{eq:lc} may be shorter than the nuclear thickness, $2R_A$, and complete longitudinal coherence may be lost.    We avoid this region by requiring $M_V^2 + Q^2 < 2\hbar k/R_A$.  This requirement restricts  near-threshold production.

\begin{table}
\begin{tabular}{|c|c|c|}
\hline
Meson & $c_1$    & $c_2$ ($10^{-2}\rm{GeV}^{-2}$)  \\
\hline
$\rho$ & $2.09\pm0.10$ & $0.73\pm 0.18$ \\
$\phi$ & $2.15 \pm 0.17$ & $0.74 \pm 0.46$ \\
$J/\psi$ & $2.36 \pm 0.20$ & $0.29 \pm 0.43$ \\
\hline
\end{tabular}
\caption{Coefficents $c_1$ and $c_2$ for VM photoproduction in Eq. \ref{eq:n}.   The values and uncertainties for $\rho$ and $\phi$ are reported in \cite{Aaron2010}. For J/$\psi$ the values were determined by us a least squares fit to HERA data.}
\label{tab:n}
\end{table}

\section{Final state generation}

To study detector acceptances, we simulate the decays of the produced VMs into their final states, accounting for the VM polarization, and simulate complete events, including the outgoing electron and proton/ion.   This requires being able to rapidly select events with a given $W$, $k$ and $Q^2$ with appropriate weighting.  

\subsection{Sampling}\label{subsec:sampling}

Because the cross-sections depend on so many kinematic variables, and are so sensitive to $Q^2$, it is impractical to calculate and then sample from a sufficiently multi-dimensional lookup table.  Instead, we refactorize and decouple the cross-section to encode the cross-section into  two-dimensional tables which can be relatively quickly generated, and then used to generate events very quickly.  To account for the rapid change in cross-section as $Q^2$ varies, we  write the full cross section as:
\begin{equation}
\frac{d^3 \sigma}{dWdkdQ^2} = n(k,Q^2)\sigma(W,Q^2)f(M_V)
\end{equation}
where $f(M_V)$ is a Breit-Wigner function for broad resonances (direct $\pi^+\pi^-$ can be included), $n(k,Q^2)$ is the photon flux discussed in \ref{eq:photon_flux} and $\sigma(W,Q^2)$ is the differential electro-nuclear cross-section discussed in section \ref{sec:photonuclear_x_sec}. For narrow resonances, $f(M_V)$ is a delta function.

These terms decouple the cross-section, a function of three variables, into three separate functions with 2 and 1 variable respectively, which are used to generate the look-up tables in the simulations.   The ranges and number of elements in the tables are matched the required kinematic range and precision.  The range in $W$ depends on the final state meson mass and width, while the range for $k$ is depends on the relevant accelerator parameters. 

Events are generated using rejection sampling.  To aid in rejection sampling, we also define a third look-up table storing the single differential photon flux:
\begin{equation}
n(k) = \int_{Q^2_{\rm min}}^{Q^2_{\rm max}} dQ^2n(k,Q^2).
\end{equation}
Each look-up table is generated during program initialization.  They are  normalized by the maximum value within the kinematic range selected for the simulation, and so are made dimensionless, in the range from 0 to 1.  A value of $M_V$ is drawn (in the case of wide resonances) along with a random variable $x_0$, and $M_V$ is accepted if $f(M_V) > x_0$. Once the finals state mass is generated, a value of $k$ is drawn along with a second random number $x_1$, and the value of $k$ is accepted if $N(k)> x_1$.  If not, another pair is drawn.

Next, the photon virtuality is selected, again using rejection sampling.  A value of $Q^2$ is drawn, along with another random number, $x_2$.  The value of $Q^2$ is accepted if $n(k,Q^2) > x_2$.  Additional pairs are generated until a pair passes this criteria.  Once a $(k,Q^2)$ pair is accepted and photon + Pomeron vertex is constructed given the invariant mass of the hadronic final state $W$. One last variable $x_3$ is drawn and the full event ($\gamma^* +X \rightarrow V+X$) is accepted if $\sigma(W,Q^2) > x_3$, otherwise the procedure starts anew.   \\
The VM $p_{T}$ is the vector sum of the photon and Pomeron $p_{T}$.  The deflection of the scattered electron ($\theta_e$) is determined by momentum transfer $Q^2$:
\begin{equation}
Q^2 = 2E_e(E_e-k)(1-\cos\theta_e).
\end{equation}
The photon $p_T$ is obtained through momentum conservation $q_{T} = (E_e-k)\sin\theta_e$.    

The momentum transfer from the nuclear targets has perpendicular and longitudinal components, $t_\perp$ and $t_{||}$, where $t_{||}$ is determined by the VM mass and longitudinal momentum conservation, so $t_{||}=M_V^2/2k$; usually $t_{||} \ll t_\perp$. 
$t_\perp$ is generated through rejection sampling following the distribution:
\begin{eqnarray}
\sigma (AA\rightarrow AAV)=\ \  &&\\ \nonumber
2 \int_0^\infty dk \frac{dN_\gamma(k)}{dk}\int_{t_{\rm min}}^\infty dt \left.\frac{d\sigma(\gamma A \rightarrow VA)}{dt}\right|_{t=0} |F(t)|^2
\end{eqnarray}
where $F(t)$ is the form factor of the target nucleus \cite{Klein:1999gv}.  The value for $t_{min} = (M_V^2/2k)^2$ is fixed for narrow resonances, but may change event-by-event for wide resonances. Here, for simplicity, $t_{min}$ is always fixed in order to explicitly decouple the $M_V$ dependence and reduce the dimensions of the look-up tables used for event generation.

For heavy ions, the form factor is taken to be the convolution of a hard sphere potential with the Yukawa potential with a range of 0.7 fm:
\begin{equation}
F(t) = \frac{4\pi\rho_0}{Aq^3}\Big[\sin(tR_{A}) - tR_{A}\cos(tR_{A})\Big]\left(\frac{1}{1+a^2t^2}\right) \ .
\end{equation}
Especially for lighter VMs, this form factor produces diffractive minima when $t$ is a multiple of $\pi/R_A$.   It should be noted that the positions of these dips may vary due to phenomena like nuclear shadowing or saturation \cite{Frankfurt:2015cwa}.   

For protons, a dipole form factor is used \cite{Drees:1989vq}
\begin{equation}
F(t) = \frac{1}{(1+q^2/0.71 {\rm GeV}^2)^2} \ .
\end{equation}
The dipole form factor predicts a featureless spectrum, decreasing with increasing $t$, in contrast to some saturation models which predict the presence of diffractive dips even with proton targets, \cite{Armesto:2014sma}. 

\begin{figure*}
\includegraphics[width = 0.7\textwidth]{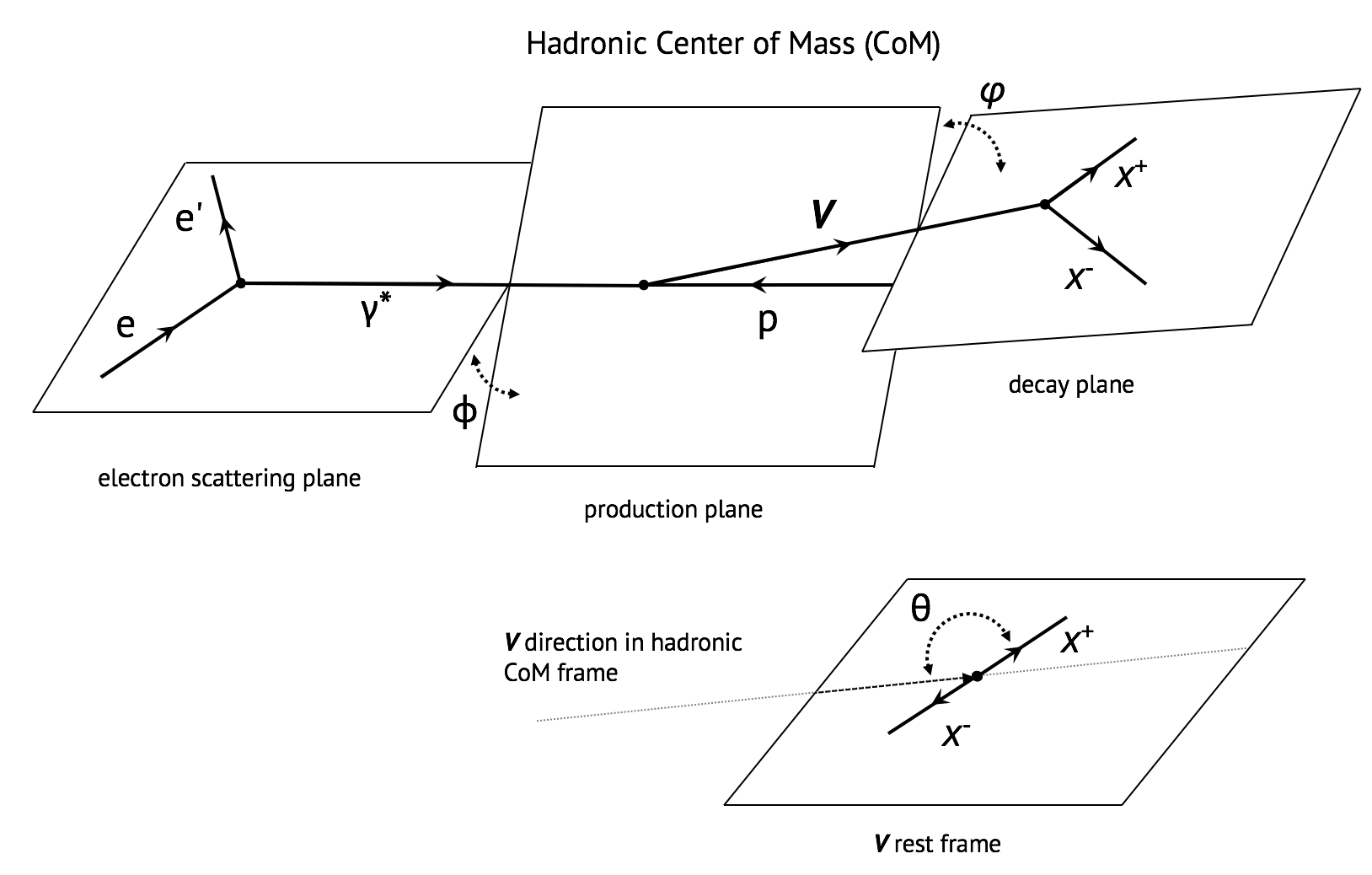}
\caption{Schematic drawing indicating the relevant angles used in describing the VM production $ep\rightarrow eVp$ and subsequent decay $V\rightarrow X^+X^-$: the angle between the electron scattering plane and the VM production plane $\phi$, the angle between the production plane and the decay plane $\varphi$ and the decay plane polar angle of the positive decay daughter $\theta$.}\label{fig:decay_schematic}
\end{figure*}

\subsection{Final state angular distributions}\label{sec:AngularDistribution}
 
Once the relevant kinematic variables have been obtained, a final random number is drawn to determine the azimuthal direction of the momentum transfer. In the limit $Q^2\rightarrow 0$, the photons are linearly polarized transverse to the beam direction; they can be decomposed into a mixture of 50\% right-handed and 50\% left-handed photons.  The VM retains the photon spin state, and the angular distributions come from the relevant spherical harmonics and Clebsch-Gordon coefficients. As the photon $Q^2$ rises, the photons can also be longitudinally polarized, along the direction of motion, and the angular distributions become more complicated.  The modifications to the the VM electroproduction cross-sections have been previously investigated, and in fact, at high enough values of $Q^2$ can be obtained perturbatively.  However, the evolution with $Q^2$ is complex, so we use the phenomenological approach described in Ref. \cite{SCHILDKNECHT1999328}.\\

We describe the VM decay in the helicity system, which  is given in terms of the three angles illustrated in Fig. \ref{fig:decay_schematic}  \cite{SCHILDKNECHT1999328,Adloffetal.2000}: 
  
\begin{itemize}
\item The angle between the electron scattering plane and VM production plane (plane defined by the VM and incoming proton) $\phi$.
\item The angle between VM production plane and the decay plane $\varphi$.
\item The polar angle, measured in the decay plane, of the positive decay product $\theta$. 
\end{itemize}

Following the same arguments, and under the approximation of $s$-channel helicity conservation (SCHC) , the longitudinal-to-transverse cross-section ratio ($R_V$) for a VM $V$ can be written as follows:

\begin{eqnarray}
R_V(W^2,Q^2) =\frac{\sigma_{L,\gamma^* p \rightarrow V p}}{\sigma_{T,\gamma^* p \rightarrow V p}} \nonumber \\
= \frac{(Q^2+m_{V,T}^2)^2}{m_{V,T}^4}\xi_V^2\left[ \frac{\pi}{2}\frac{m_{V,L}^2}{Q^2} - \frac{m_{V,L}^3}{\sqrt{Q^2}(Q^2+m_{V,L}^2)}\right. \nonumber\\
\left. - \frac{m_{V,L}^2}{Q^2}\arctan\big(\frac{m_{V,L}}{\sqrt{Q^2}}\big)\right]^2
\label{eq:ratioparametrization}
\end{eqnarray}

This expression has three parameters.  $m_{V,L}$ and $m_{V,T}$ are modified pole masses, and $\xi_V$ is the ratio of the imaginary forward scattering cross-sections for longitudinally and transversely polarized VMs.  These parameters were determined in \cite{SCHILDKNECHT1999328}, by fitting to HERA data for the $\rho$ and $\phi$ mesons:

\begin{eqnarray}
\xi_\rho = 1.06 , \quad m_{\rho,T}^2 = 0.68 m_\rho^2  , \quad  m_{\rho,L}^2 = 0.71 m_\rho^2 \nonumber \\
\xi_\phi = 0.9 ,\ \quad m_{\phi,T}^2 = 0.41 m_\phi^2  , \quad  m_{\phi,L}^2 = 0.57 m_\phi^2 \nonumber \\
\end{eqnarray}

These fits are successful at describing the HERA data, but similar results are not available for the heavier VMs which are of particular interests for an EIC. So, we make the the following simplification:
\begin{eqnarray}
\xi_V \rightarrow 1, \quad m^2_{V,L} = m^2_{V,T}  \rightarrow 0.6M_V^2
\end{eqnarray}
where $M_V$ is the actual mass of the VM.\\

The ratios $R_V$ are of particular interest since they can be related to elements of the VM spin matrix $r$ by the following expression:
\begin{equation}
R = \frac{1}{\epsilon}\frac{r^{04}_{00}}{1-r^{04}_{00}}
\label{eq:polratio}
 \end{equation}
where $\epsilon$ is the polarization parameter of the virtual photon, itself given by:   
\begin{equation}
\epsilon \simeq \frac{1-y}{1-y+y^2/2}, \qquad y = \frac{P\cdot q}{P\cdot p} = 1-\frac{E_e'}{E_e}\cos^2\left(\frac{\theta}{2}\right)
\label{eq:polpara}
\end{equation}

\begin{figure*}
\center
\subfloat[][]{
	\includegraphics[width=0.45\textwidth]{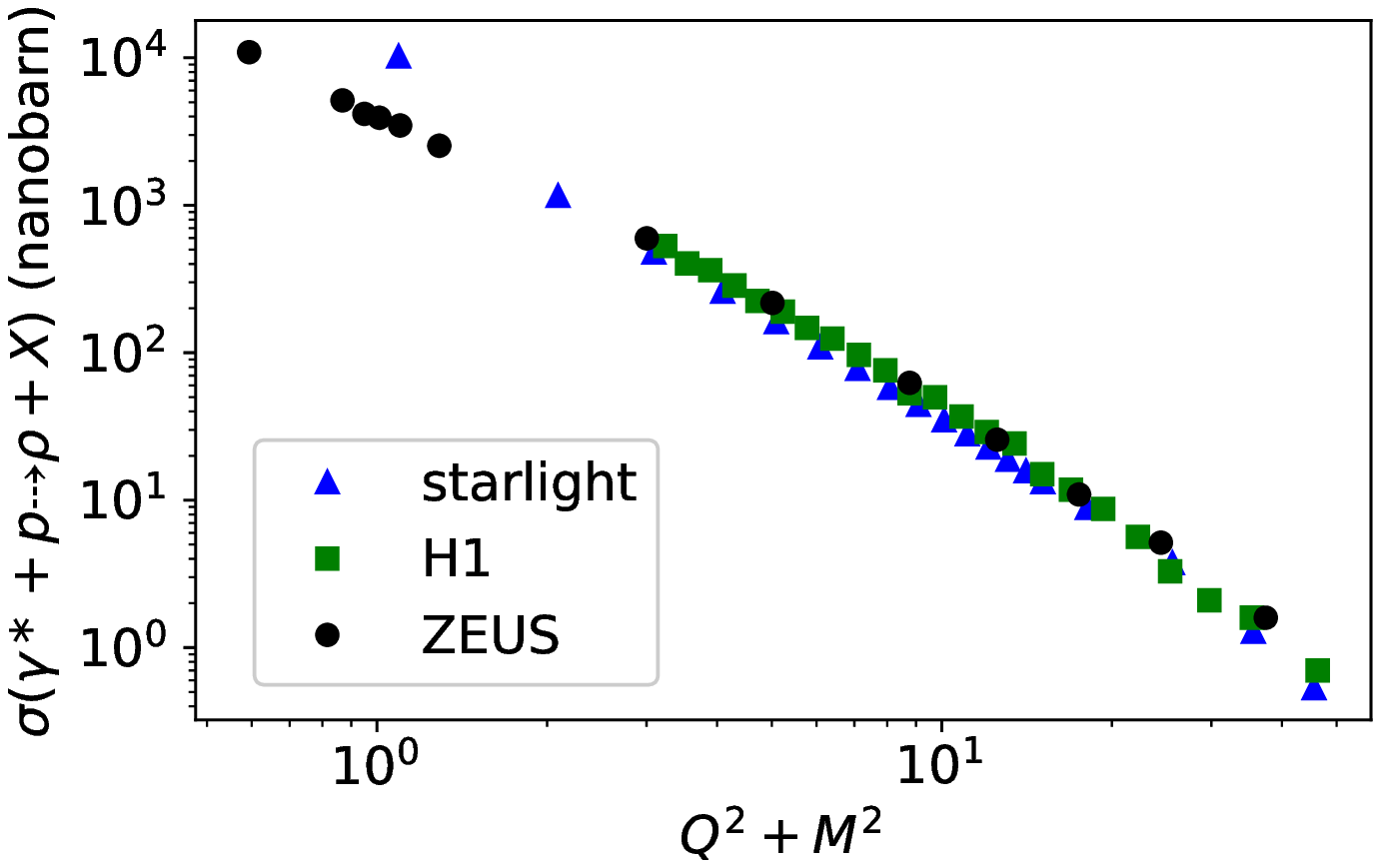}}
\subfloat[][]{
	\includegraphics[width=0.45\textwidth]{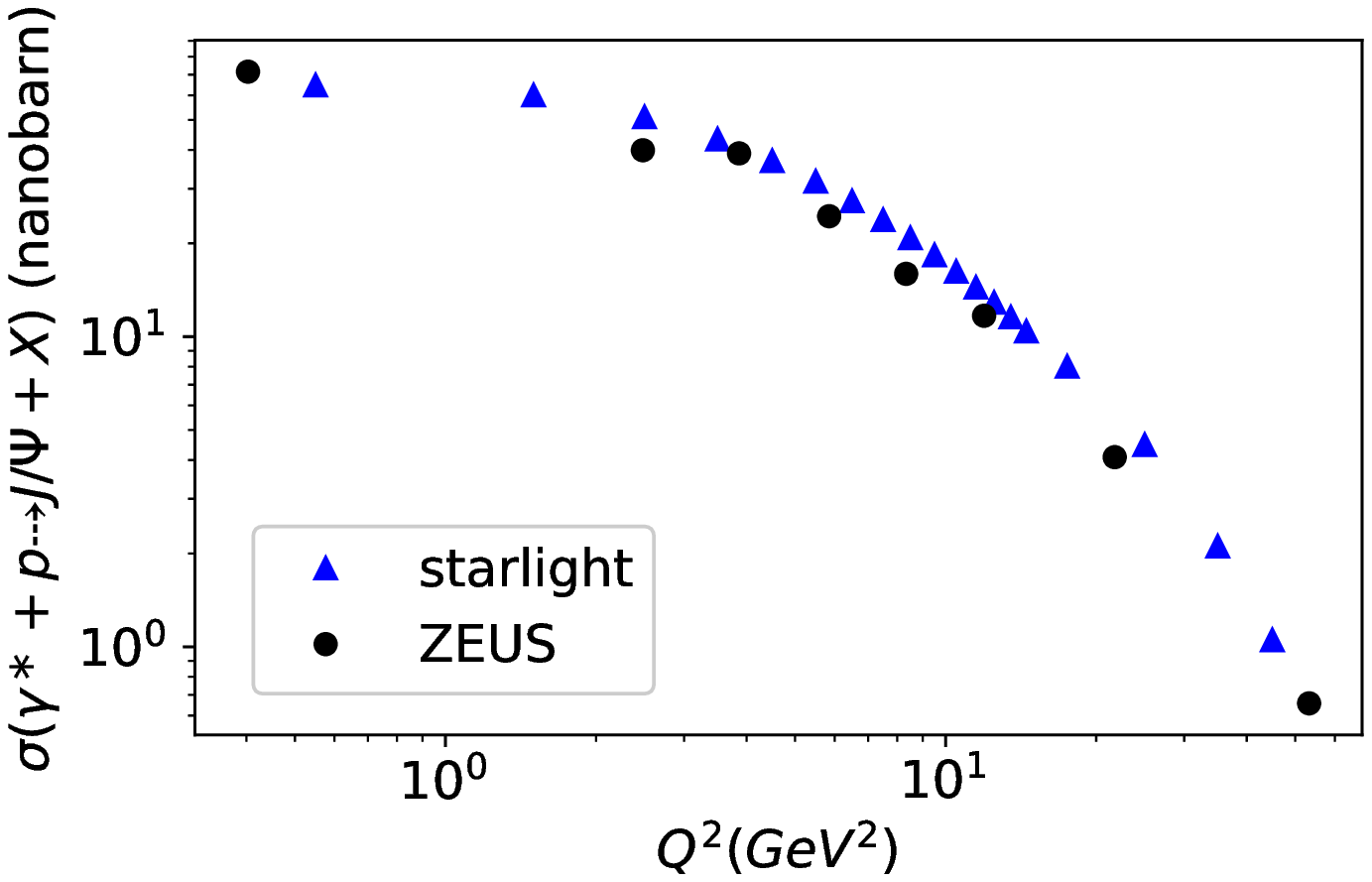}}\\
\subfloat[][]{
	\includegraphics[width=0.45\textwidth]{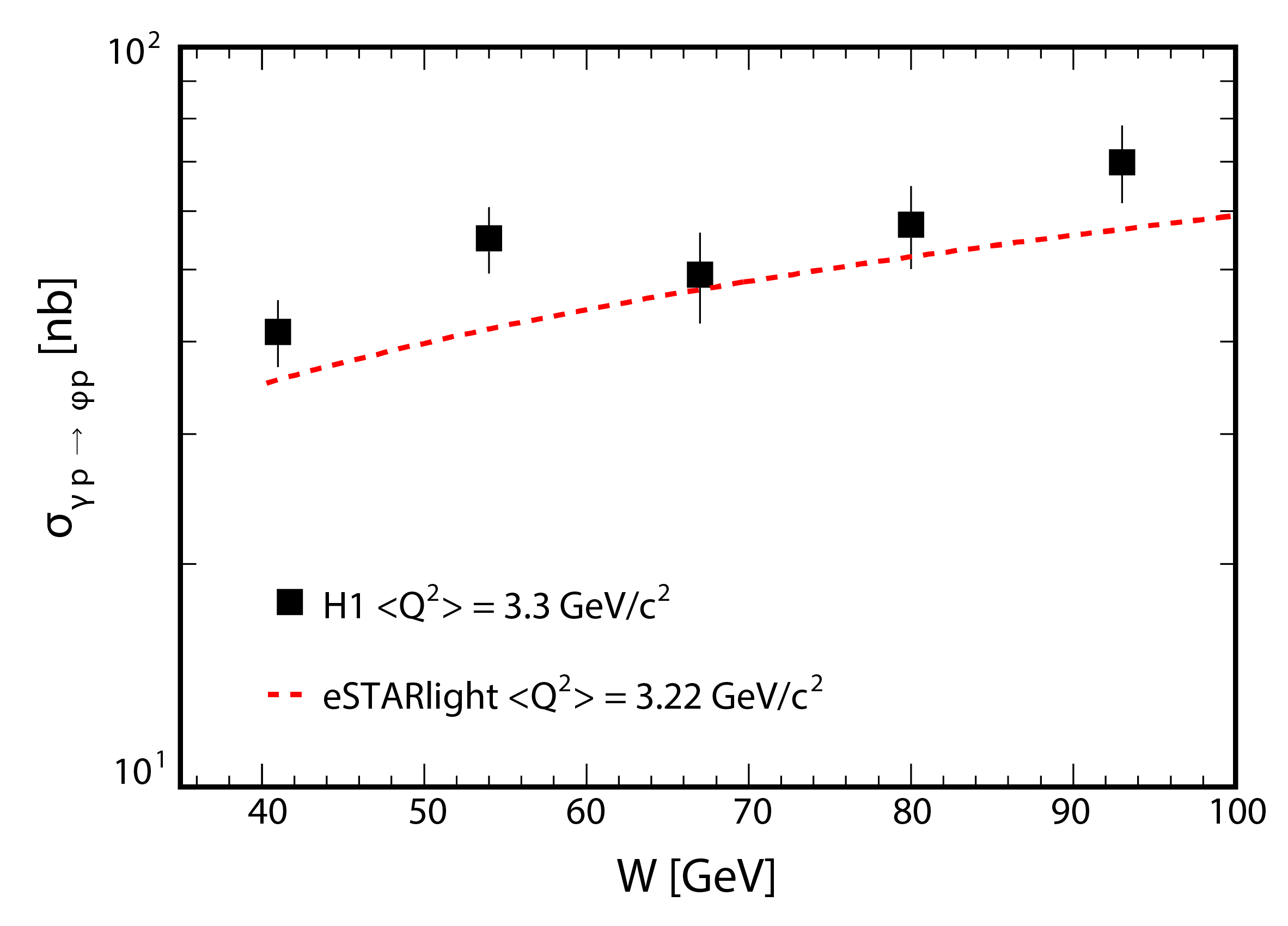}}
\caption{Calculations of photo-nuclear cross sections $\sigma(\gamma p \rightarrow Vp)$ vs. $Q^2$ (panels a and b) and $W$ (panel c) compared with HERA data.  The $x$ axes for the panels a and b are chosen to match that from the experimental data shown.  The data is from Refs. \cite{Aaron2010} and \cite{Chekanov20043} respectively. \label{fig:HERA} }
\label{fig:hera_vs_q2}
\end{figure*}
   
In the SCHC limit the angle $\phi$ is described by a uniform distribution. Although previous studies have shown that this can violated by as much as 10\% in electroproduction of VMs \cite{SCHILDKNECHT1999328}, there is no consistent treatment for all VM species. So, we assume SCHC holds, and take the single differential distributions (integrated over $\phi$ and $\varphi$ or $\theta$ respectively) to sample during event generation. For spin-0 VM, such as $\rho$ and $\phi$, the angular distributions are, after integration \cite{Adloffetal.2000}:
\begin{eqnarray}
\Omega(\cos\theta) &\propto& 1 - r_{00}^{04} + (3r_{00}^{04}-1)\cos^2(\theta) \\
\Omega(\cos\varphi) &\propto& 1 -2r^{04}_{1-1}\cos(2\varphi)
\end{eqnarray}
while for the spin-1/2 states, similar expressions can be obtained \cite{Breitwegetal.1999}:

\begin{eqnarray}
  \Omega(\cos\theta) &\propto& 1 + r_{00}^{04} + (1-3r_{00}^{04})\cos^2(\theta) \\
  \Omega(\cos\varphi) &\propto& 1 +r^{04}_{1-1}\cos(2\varphi)
\end{eqnarray}

In both cases, fits to the data yield a value for $r_{1-1}^{04}$ that is consistent with zero, so we take it to be zero, with the angle $\varphi$ uniformly distributed in the range $[0,2\pi]$. The angle $\theta$ is generated once the kinematics of the event have been determined, obtaining the value of spin matrix element $r_{00}^{04}$ from the photon $Q^2$ and using the phenomenological fit in Eq. \ref{eq:ratioparametrization} together with Eqs. \ref{eq:polratio} and \ref{eq:polpara}.
  
\section{Results: Cross-sections and Kinematic Distributions}

In this section, we present calculations for the production of a variety of mesons in $ep$ and $eA$ collisions.    We consider two ranges of $Q^2$: $Q^2_{\rm min} < Q^2 < 1$ GeV$^2$ (photoproduction, PP) , and $Q^2 > 1$ GeV$^2$ (electroproduction, EP).  The 1 GeV$^2$ dividing line is chosen following Chapter 3 of Ref. \cite{Accardi:2012qut}.  

As a check, we compared our results with observations from the HERA collider, a sample of which are shown in Figure \ref{fig:hera_vs_q2}. Panels (a) and (b) show the photo-nuclear cross sections $\sigma( \gamma p \rightarrow V p )$ vs $Q^2$ for $\rho$ and $J/\psi$ compared to the experimental data obtained at HERA by the the H1 and ZEUS experiments \cite{Aaron2010,Chekanov20043}.  The third panel (c) shows the photo-nuclear cross section for $\phi$ production versus the invariant mass of the hadronic system ($W$) for a single $Q^2$ bin.

\begin{table*}
\begin{tabular}{|l|r|r|r|r|r|r|r|r|r|r|}
\hline
Accelerator 	& \multicolumn{5}{c|}{$\sigma$} & \multicolumn{5}{c|}{Number of events} \\
			& $\rho^0$ & $\phi$      & J$/\psi$    & $\psi'$     & $\Upsilon(1S)$
			& $\rho^0$ & $\phi$      & J$/\psi$    & $\psi'$     & $\Upsilon(1S)$ \\	
\hline
eRHIC - $ep$ & $5.0$ $\mu$b & $230.0$ nb & $8.5$ nb & $1.4$ nb & $14.0$ pb & 50 giga & 2.3 giga & 85 mega & 14 mega & 140 kilo\\
eRHIC - $eA$ & $870.0$ $\mu$b & $55.0$ $\mu$b & $1.9$ $\mu$b & $320.0$ nb & $1.2$ nb & 44 giga & 2.8 giga & 100 mega & 16 mega & 60 kilo\\
\hline
JLEIC - $ep$ & $3.7$ $\mu$b & $160.0$ nb & $3.9$ nb & $600.0$ pb & $4.3$ pb & 37 giga & 1.6 giga & 39 mega & 6.0 mega & 43 kilo\\
JLEIC - $eA$ & $580.0$ $\mu$b & $33.0$ $\mu$b & $590.0$ nb & $82.0$ nb & - & 28 giga & 1.6 giga & 28 mega& 3.9 mega & -\\
\hline
LHeC - $ep$ & $10.0$ $\mu$b & $560.0$ nb & $47.0$ nb & $7.8$ nb & $120.0$ pb & 100 giga & 5.6 giga & 470 mega & 78 mega & 1.2 mega\\
LHeC - $eA$ & $2.3$ mb & $170.0$ $\mu$b & $15.0$ $\mu$b & $2.9$ $\mu$b & $41.0$ nb & 110 giga & 8.2 giga & 720 mega & 140 mega & 2.0 mega\\
\hline
HERA - $ep$ & $7.9$ $\mu$b & $450.0$ nb & $40.0$ nb & $6.4$ nb & $85.0$ pb & - & - & - & - & -\\
\hline
\end{tabular}
\caption{The cross-sections and rates for VM photoproduction ($Q^2<1$ GeV$^2$) at the proposed EICs, and at HERA.
\label{tab:photo_prod} }
\end{table*}

\begin{table*}
\begin{tabular}{|l|r|r|r|r|r|r|r|r|r|r|}
\hline
Accelerator 	& \multicolumn{5}{c|}{$\sigma$} & \multicolumn{5}{c|}{Number of events} \\
			& $\rho^0$ & $\phi$      & J$/\psi$    & $\psi'$     & $\Upsilon(1S)$
			& $\rho^0$ & $\phi$      & J$/\psi$    & $\psi'$     & $\Upsilon(1S)$ \\	
\hline
eRHIC - $ep$ & $14.0$ nb & $1.7$ nb & $570.0$ pb & $120.0$ pb & $2.4$ pb & 140 mega & 17 mega & 5.7 mega & 1.2 mega & 24 kilo\\
eRHIC - $eA$ & $730.0$ nb & $110.0$ nb & $77.0$ nb & $19.0$ nb & $200.0$ pb & 37 mega & 5.6 mega & 3.9 mega & 960 kilo & 10 kilo\\
\hline
JLEIC - $ep$ & $10.0$ nb & $1.2$ nb & $270.0$ pb & $55.0$ pb & $790.0$ fb & 100.0 mega & 12 mega & 2.7 mega & 550 kilo & 7.9 kilo\\
JLEIC - $eA$ & $450.0$ nb & $67.0$ nb & $25.0$ nb & $5.1$ nb & - & 22 mega & 3.2 mega & 1.2 mega & 250 kilo & -\\
\hline
LHeC - $ep$ & $26.0$ nb & $3.7$ nb & $2.9$ nb & $630.0$ pb & $18.0$ pb & 260 mega & 37 mega & 29 mega & 6.3 mega & 180 kilo\\
LHeC - $eA$ & $2.0$ $\mu$b & $340.0$ nb & $560.0$ nb & $150.0$ nb & $5.3$ nb & 100 mega & 16 mega & 27 mega & 7.2 mega & 250 kilo\\
\hline
HERA - $ep$ & $44.0$ nb & $6.4$ nb & $17.0$ nb & $3.6$ nb & $120.0$ pb & - & -& -& - & -\\
\hline
\end{tabular}
\caption{The cross-sections and rates for VM electroproduction ($Q^2 > 1$ GeV$^2$) at the proposed EICs and at HERA.
\label{tab:electro_prod} }
\end{table*}

Tables \ref{tab:photo_prod} and \ref{tab:electro_prod} show, respectively, the photo-production and electroproduction $ \sigma(eX\rightarrow eVX) $ cross sections and expected rates for different species at the different accelerators.  The rate calculations are for an integrated luminosity of 10 fb$^{-1}$ for protons, and 10 fb$^{-1}$/nucleon for heavy ions.   The much larger cross-sections for ions are mostly offset by the decreased per-ion lumosity, leading to only slightly higher rates for $eA$ collisions than for $ep$.   The use of a quantum Glauber calculation and neglect of shadowing could lead to some over-estimation in the cross-sections for ion targets. 

The electroproduction rates are roughly 100 times smaller than the corresponding photoproduction rates; this is a consequence of the lower photon flux as $Q^2$ rises, coupled with the decrease in the photon-nucleon cross-section with increasing $Q^2$.  

\begin{figure}
\subfloat[][]{
	\includegraphics[width=0.45\textwidth]{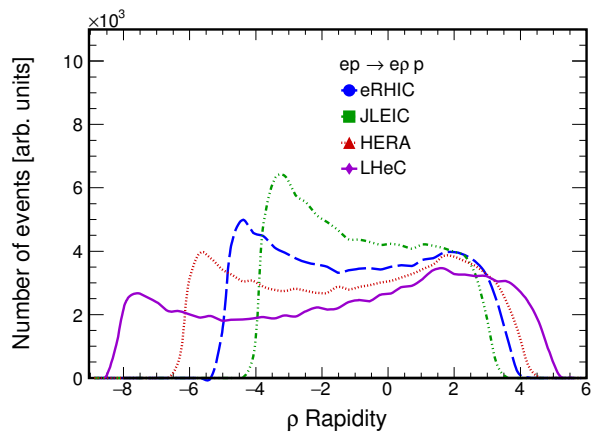}}\\
\subfloat[][]{
	\includegraphics[width=0.45\textwidth]{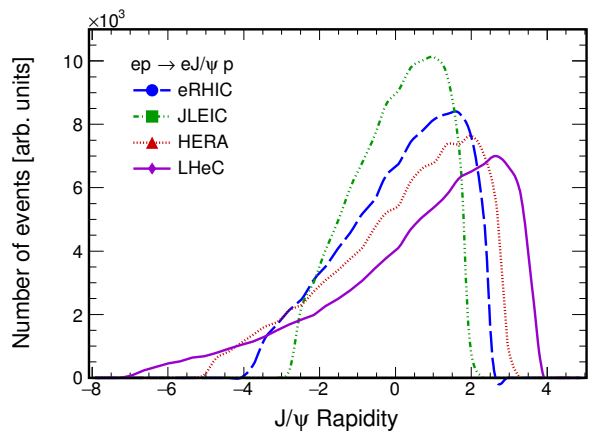}}
\caption{Predictions for $dN/dy$ for a) $\rho$ and b) $J/\psi$ meson production in $ep$ collisions at the different proposed facilities.}
\label{fig:dndy}
\end{figure}

Figure \ref{fig:dndy} shows the predicted rapidity distribution for $\rho$ (top) and $J/\psi$ (bottom) production in $ep$ collisions at the different proposed colliders, and at HERA.   The rapidity is related to the photon energy; neglecting transverse momentum, 
\begin{equation}
k=M_V/2 \exp(y).
\end{equation}
This is in the lab frame; in the target frame $k=2\gamma_I M_V/2 \exp(y)$, where $\gamma_I$ is the Lorentz boost of the ion.    Large negative rapidity corresponds to low photon energy, while large positive rapidity corresponds to high photon energy.  The rapidity may also be used to find the Bjorken-$x$ of the struck parton: $x=M_V/(2\gamma_I m_p) \exp(-y)$, where $m_p$ is the proton mass.  To probe low-$x$ gluons, it is necessary to go to forward rapidity. 

For the $\rho$, two peaks are visible in the distributions.  The negative rapidity (low-energy $k$) peak corresponds mostly to production via meson exchange near threshold, while the positive rapidity (higher $k$) peak is from Pomeron exchange.  For the $J/\psi$, only the second peak is present; it is sharper than for the $\rho$ because of the faster increase in cross-section with increasing $k$.   Meson exchange contributes a significant fraction of the cross-section for $\rho$ and  $\phi$ and their excited states; one cannot neglect it.  

\begin{figure}
\subfloat[][]{
	\includegraphics[width=0.45\textwidth]{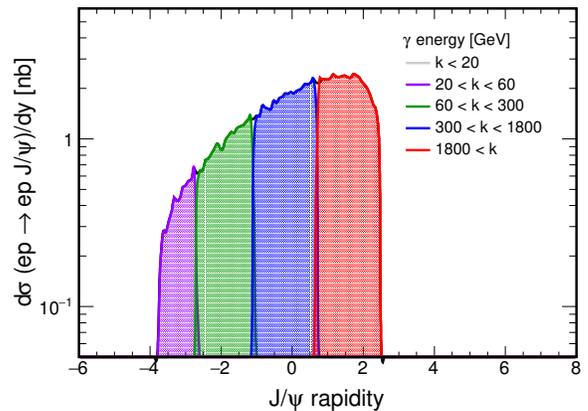}}\\
\subfloat[][]{
	\includegraphics[width=0.45\textwidth]{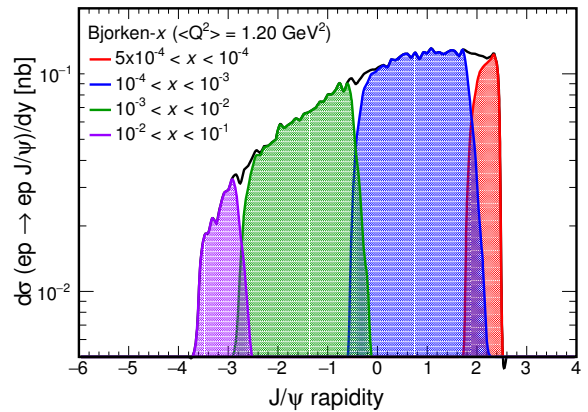}}
\caption{$d\sigma/dy$ for production of $J/\psi$ in $ep$ collisions at the proposed eRHIC accelerator, a) divided up by photon energy (in the target frame) into five bands.  Panel b) shows $J/\psi$ production for $1 < Q^2 < 10$ $\rm{GeV}^2$ divided into  four bands according to the Bjorken-$x$ of the struck parton.}
\label{fig:dndyvsk}
\end{figure}

Fig. \ref{fig:dndyvsk}  shows the rapidity range for $\rho$ and $J/\psi$ production for $ep$ collisions at the eRHIC collider, divided up into five photon energy bands, showing the relationship between photon energy and rapidity.  It is immediately apparent that studying a wide range of photon energies requires a detector with large acceptance in rapidity; scanning in rapidity is roughly equivalent to scanning in $1/x$.  One could probe lower photon energies by running the accelerator at a reduced collision energy, but the only way to study the highest energy photons is with a forward detector. 

\begin{figure}
\subfloat[][]{
\includegraphics[width=0.45\textwidth]{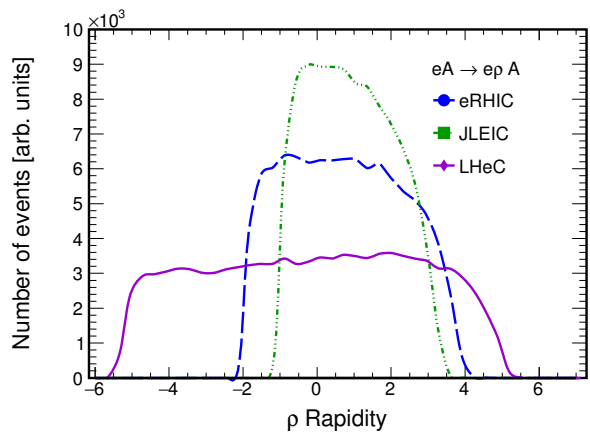}}\\
\subfloat[][]{
\includegraphics[width=0.45\textwidth]{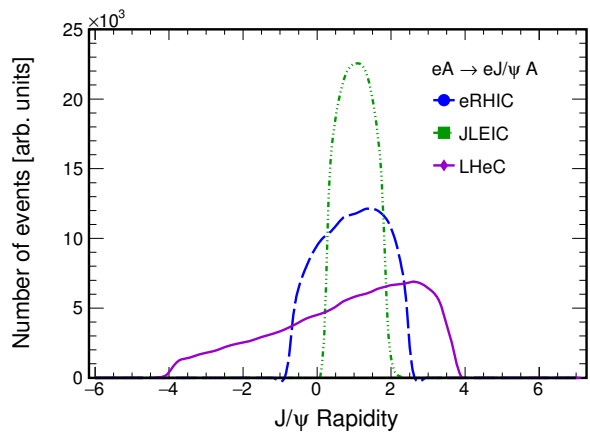}}
\caption{Predictions for $dN/dy$ for photoproduction $eA\rightarrow e V A$ for a) $\rho$ and b) $J/\psi$ at the different proposed colliders.}
\label{fig:dndy_eA}
\end{figure}

Figure \ref{fig:dndy_eA} shows predictions for $\rho$ and $J/\psi$ production in $eA$ collisions at the different proposed colliders. The kinematic range available for VM production is reduced due to the smaller center of mass energy per nucleon in the system, leading to production over a smaller range in rapidities. The coherence requirement limits $t$ for larger ions, reducing production near the photon-nucleon energy threshold, at large negative rapidities. 

\subsection{Electroproduction}

\begin{figure}
\center
\subfloat[][]{
	\includegraphics[width=0.45\textwidth]{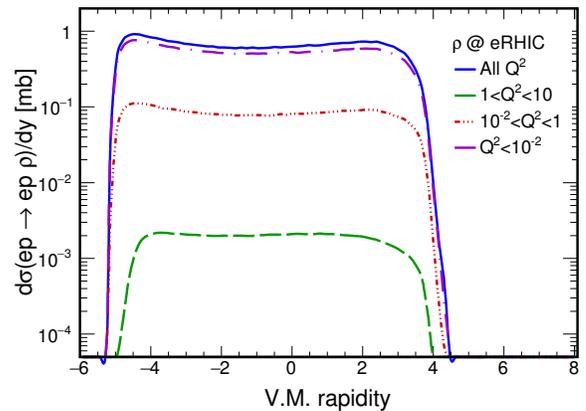}}\\
\subfloat[][]{
	\includegraphics[width=0.45\textwidth]{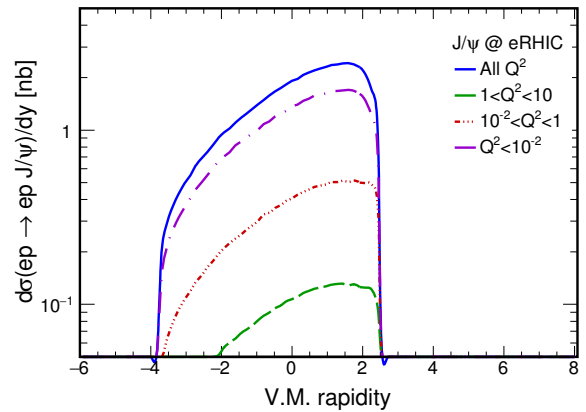}}
\caption{Predictions for $d\sigma/dy$ for electroproduction of a) $\rho$ and b) $J/\psi$ in $ep$ collisions at the eRHIC for all $Q^{2}$ and three different bins: $Q^2 < 10^{-2} \rm{ GeV}^2$, $10^{-2} < Q^2 < 1 \rm{GeV}^2$ and $ 1 < Q^2 < 10 \rm{GeV}^2$.}
\label{fig:dndy_q2_gt_2}
\end{figure}

Figure \ref{fig:dndy_q2_gt_2} shows predictions for $d\sigma/dy$ for a) $\rho$ and b) $J/\Psi$ production in $ep$ collisions at eRHIC for four different $Q^2$ ranges.   The distribution narrows with increasing $Q^2$.    


\begin{figure}
\includegraphics[width=0.45\textwidth]{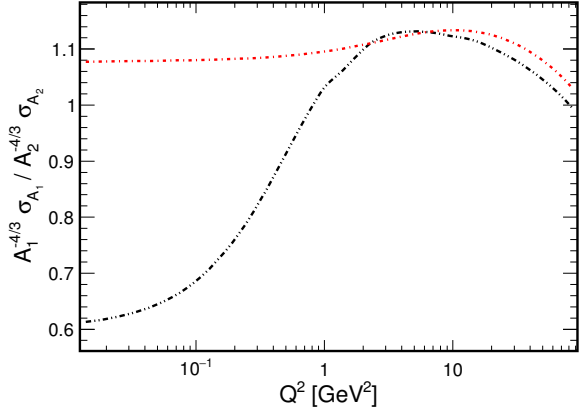}
\caption{The scaled (by $A^{-4/3}$) ratio of the $\rho$ (black curve) and $J/\psi$ (red) photoproduction cross-section on lead and iron targets, as a function of $Q^2$.  Without nuclear effects, this ratio should be 1.  At low $Q^2$, the $\rho$ ratio is much less than one because of nuclear shadowing.  As $Q^2$ rises,  the nucleus becomes much more transparent, and the scaled ratio rises to 1.  For the $J/\psi$ $\sigma_{\gamma^*A\rightarrow VA}$ is always relatively small, so the nuclear suppression at low $Q^2$is much smaller.}
\label{fig:dndyscaled}
\end{figure}

 Figure \ref{fig:dndyscaled} shows the evolution of the $\rho$ and $J/\psi$ cross-sections with $Q^2$.   Following Ref. \cite{Mantysaari:2017slo}  we plot the ratio of the cross-section on gold and iron targets, scaled by $A^{-4/3}$.  This scaling is chosen because, in the absence of nuclear effects, the integrand in Eq. (8) reduces to $\sigma_{\rm tot}(Vp)T_{AA}(\vec{r})$ and the coherent forward scattering cross-section scales as $A^2$.  The maximum momentum transfer decreases as $\hbar/R_A \approx A^{-1/3}$.  The transverse phase space decreases as $A^{-2/3}$, leading to an overall coherent cross-section that scales as $A^{4/3}$.   So, without any nuclear effects, the scaled ratio should be 1.  When nuclear shadowing is occurs, the ratio will drop.  
 
For the $J/\psi$, and for the $\rho$ for $Q^2 \gtrapprox 5$ GeV$^2$, the ratio is slightly greater than 1.  In this regime, multiple interactions by a single dipole are not important.  The rise in ratio slightly above 1 is likely due to the coherence condition, Eq. \ref{eq:lc}, which allows a larger momentum transfer $t$ from the iron nuclei than from gold.   For the $\rho$, as $Q^2$ decreases, the ratio drops to about 0.62, reflecting a strong nuclear shadowing effect.  For the $J/\psi$, nuclear shadowing is visible, but it is much smaller than for the $\rho$.   Charmed quarks are heavy, so the $c\overline c$ dipole is always small.    Both ratios drop at very large $Q^2$, near the kinematic limits; these are reached for gold before iron, because the smaller nuclear size allows a higher maximum photon energy.  These trends are qualitatively similar to those shown in Fig. 2 of Ref. \cite{Mantysaari:2017slo}, which is based on a model including nuclear saturation. 

So, the study of $J/\psi$ electroproduction at large $Q^2$ requires instrumentation at large negative rapidity.  This conclusion also applies to the production of other heavy mesons, where the cross-section rises rapidly with increasing $k$. 
  
\subsection{The outgoing electrons and ions}

\begin{figure}[b]
	\includegraphics[width=0.45\textwidth]{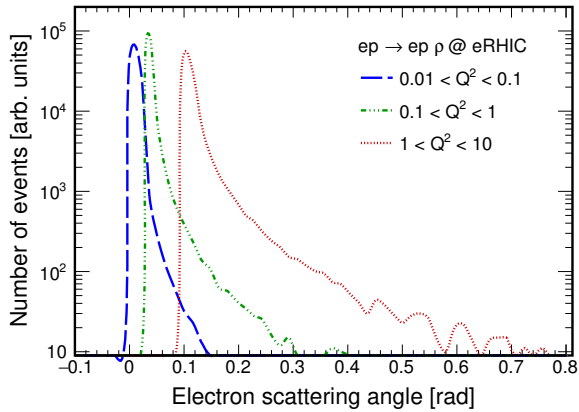}
\caption{Electron scattering angle for $J/\psi$ production at eRHIC for three $Q^2$ ranges. The scattering angle also depends on the photon energy and vector meson mass.}
\label{fig:scattered_e}
\end{figure}

At an EIC, detection of the scattered electron is important since it provides information on the photon kinematics.    In most photoproduction reactions, the electron loses energy but undergoes little deflection; as $Q^2$ rises, so does the deflection angle.  Figure \ref{fig:scattered_e} shows the rapidity of the scattered electron for coherent $\rho$ production in three different $Q^2$ bins.
 The lower limit of  $(0.1\rm{GeV})^{2}$ is below our definition of electroproduction, but illustrates what happens at low $Q^2$.  For photoproduction, at still smaller $Q^2$, the deflection angle decreases further, and a small-angle tracker would be required to observe the scattered electrons.  This is highly desirable, since either the final state electron or ion must be detected in able to fully constrain the event kinematics. 

\begin{figure}
	\includegraphics[width=0.45\textwidth]{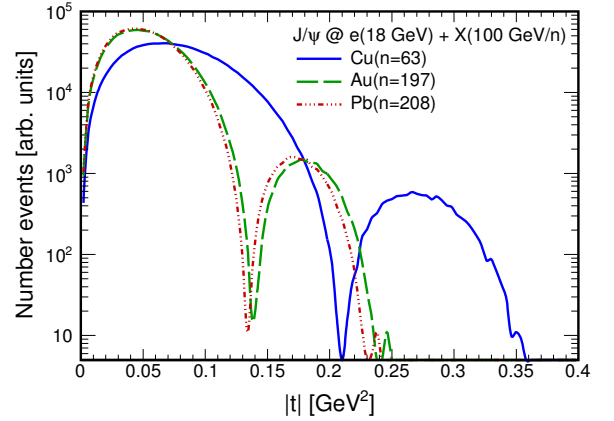}
\caption{Predictions for $dN/dt$ for photoproduction $eX\rightarrow eJ/\psi X$ at eRHIC for three different nuclear targets: copper, gold and lead.}
\label{fig:dsdt}
\end{figure}

The momentum transfer $t$ from the proton/ion is also of interest, since it is critical in probing the structure of the target.  As long as the outgoing electron is observed, $t$ can be reconstructed from the VM and electron kinematics.    Figure \ref{fig:dsdt} shows $dN/dt$ for generated $J/\psi$ events in $eA$ collisions at eRHIC energies for three different nuclear targets (at 100 GeV/n).  Because of the complete kinematic determination, deep diffraction minima are visible.  This contrasts with ultra-peripheral collisions, where the photon $p_T$ partially fills in these minima \cite{Klein:1999gv,Adamczyk:2017vfu}.    In the model used here, the diffractive minima depend only on the size of the target nucleus; saturation effects \cite{Accardi:2012qut} or nuclear shadowing \cite{Guzey:2016qwo} can shift the positions of these minima.

Even though $t$ can be reconstructed using the outgoing electron and VM kinematics, there are still reasons to observe the outgoing ion directly.  The main reason is to check if the outgoing ion remains intact, or was dissociated or excited.  Many excitations lead to observable changes in the kinematics, but nuclear dissociation has a much smaller effect on the rest of the event.  Outgoing protons may be detectable with Roman pots, as has been done with for the RHIC proton beams \cite{Bultmann:2004ke}.  However, for heavy ions, only proton or neutron fragments are likely to be visible, providing only an indication if the ion dissociated or not.  Detectors that use zero-degree calorimeters need to account for  additional nuclear dissociation events, which can produce neutrons which may overlap with VM photoproduction where the target remains intact \cite{Klein:2014xoa}. 

\subsection{Final state particle distributions and toy detector acceptance.}

Most VMs have simple final states - two charged particles -, which, along with the outgoing electron and proton or ion, constitute the entire final state.   The observability of the final state particles depends on the detector acceptance in pseudorapidity, $\eta$, and transverse momenta, $p_T$.  We assume that the eventual EIC detectors will have good coverage down to low $p_T$, so focus on the $\eta$ dependence.   

\begin{figure}[h]
\includegraphics[width=0.45\textwidth]{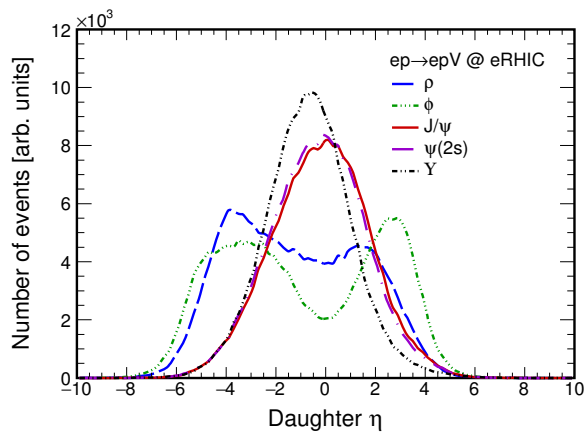}
\caption{Predictions for the final state decay products pseudorapidity for $\rho\rightarrow\pi\pi$, $\phi\rightarrow KK$, $J/\psi\rightarrow ee$, $\psi(2s)\rightarrow ee$ and $\Upsilon\rightarrow ee$ at eRHIC.    The shapes of the $\rho$ and $\phi$  final states are very different from the heavier VMs, due to the different angular distributions for VMs to decay to two spin 0 mesons, compared to the angular distribution for two spin 1/2 leptons.
}
\label{fig:final_state_eta}
\end{figure}

Fig. \ref{fig:final_state_eta} shows the $\eta$ distribution for final state decays from $\rho\rightarrow\pi\pi$,  $\phi\rightarrow KK$, $J/\psi\rightarrow ee$, $\psi(2s)\rightarrow ee$ and $\Upsilon\rightarrow ee$.    A double-hump structure is visible for VMs that decay to two spin-0 mesons: the $\rho$ and $\phi$, and absent for the final states that decay to lepton pairs.  This is a consequence of the different angular distributions in Eqs. 23 and 24, for the two classes of decays \cite{Klein:2016yzr}. 

\begin{figure}[h]
\includegraphics[width=0.45\textwidth]{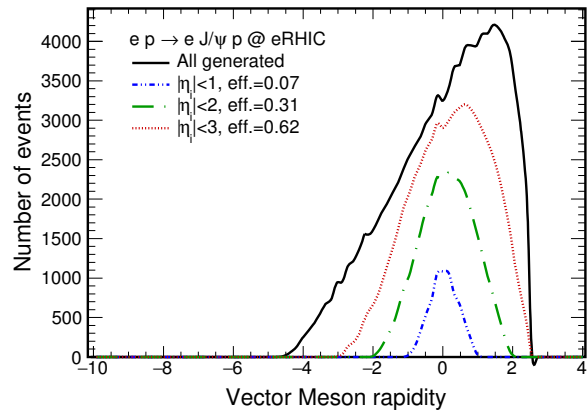}
\includegraphics[width=0.45\textwidth]{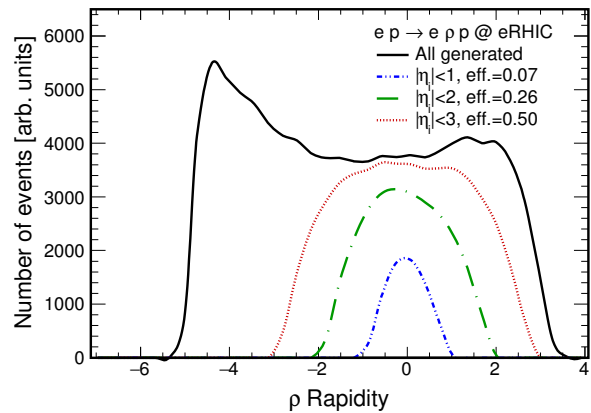}
\caption{( The acceptance for (top) $J/\psi$ and (bottom) $\rho$ production in $ep$ collisions at eRHIC for detectors with different ranges of rapidity acceptance.
}
\label{fig:VM_acceptance}
\end{figure}
 
The middle panel in Fig. \ref{fig:VM_acceptance} compares all generated $J/\psi$ events in $ep$ collisions at eRHIC with final states where both daughters would be seen in a central detector covering $\pm$1, $\pm$2 or $\pm$3 units of pseudorapidity.  The acceptance drops rapidly with decreasing detector acceptance.    By doubling the coverage ($|\eta|<1\rightarrow 2$) we can achieve the same statistical significance with a quarter of the number collisions.  More importantly, the limited acceptance reduces the range of photon energies  and Bjorken-$x$ values that can be studied, as can be seen in Fig. \ref{fig:dndyvsk}.

\section{Discussion and Conclusions}

We have calculated the kinematics for photoproduction and electroproduction at an electron-ion collider for a variety of VMs.    The kinematic distributions are important in determining the requirements for EIC detectors.  Although the details will depend on the specific accelerator configuration, some general conclusions may be drawn.

First, rates are very high for light mesons, and even for the $J/\psi$.  For eRHIC and JLEIC, $\Upsilon$ photoproduction will produce a useful sample, but the rates for electroproduction are moderate, inadequate for detailed studies with multi-dimensional binning.  At LHeC, the rates for all mesons are large. 

Second, VM photoproduction occurs at a wide range of rapidities, corresponding to the photon energy.  To study production at low photon energies (overlapping with the current studies at JLab), requires a detector that is sensitive at large negative rapidities, while studies of production at high photon energies, near the kinematic limits, requires a detector that is sensitive at large positive rapidities.   

Third, for photoproduction, the electron deflection angles are very low, particularly for low-energy photons.   A very small angle tagging system will be necessary to completely reconstruct these events.   For electroproduction, the angles are larger.

Fourth, the pseudorapidity distribution of the particles from the vector-meson decay depends both on the VM rapidity and on the spin structure of the decay.  Pions or kaons from VM decays to two spin-0 mesons are more widely spread in rapidity than dilepton final states. 

 The need for a large acceptance detector may be partially alleviated by taking data at multiple collision energies.  With a lower electron energy, one can shift the VM final states toward mid-rapidity, and also increase the deflection angle of the outgoing electron. 

We thank Joakim Nystrand, Janet Seger, Joey Butterworth and Yury Gorbunov for their work with STARlight, which provided the code base to start this work.   This work was funded by the U.S. Department of Energy under contract number DE-AC-76SF00098.

 \bibliography{paper_biblio}

\begin{thebibliography}{45}%
\makeatletter
\providecommand \@ifxundefined [1]{%
 \@ifx{#1\undefined}
}%
\providecommand \@ifnum [1]{%
 \ifnum #1\expandafter \@firstoftwo
 \else \expandafter \@secondoftwo
 \fi
}%
\providecommand \@ifx [1]{%
 \ifx #1\expandafter \@firstoftwo
 \else \expandafter \@secondoftwo
 \fi
}%
\providecommand \natexlab [1]{#1}%
\providecommand \enquote  [1]{``#1''}%
\providecommand \bibnamefont  [1]{#1}%
\providecommand \bibfnamefont [1]{#1}%
\providecommand \citenamefont [1]{#1}%
\providecommand \href@noop [0]{\@secondoftwo}%
\providecommand \href [0]{\begingroup \@sanitize@url \@href}%
\providecommand \@href[1]{\@@startlink{#1}\@@href}%
\providecommand \@@href[1]{\endgroup#1\@@endlink}%
\providecommand \@sanitize@url [0]{\catcode `\\12\catcode `\$12\catcode
  `\&12\catcode `\#12\catcode `\^12\catcode `\_12\catcode `\%12\relax}%
\providecommand \@@startlink[1]{}%
\providecommand \@@endlink[0]{}%
\providecommand \url  [0]{\begingroup\@sanitize@url \@url }%
\providecommand \@url [1]{\endgroup\@href {#1}{\urlprefix }}%
\providecommand \urlprefix  [0]{URL }%
\providecommand \Eprint [0]{\href }%
\providecommand \doibase [0]{http://dx.doi.org/}%
\providecommand \selectlanguage [0]{\@gobble}%
\providecommand \bibinfo  [0]{\@secondoftwo}%
\providecommand \bibfield  [0]{\@secondoftwo}%
\providecommand \translation [1]{[#1]}%
\providecommand \BibitemOpen [0]{}%
\providecommand \bibitemStop [0]{}%
\providecommand \bibitemNoStop [0]{.\EOS\space}%
\providecommand \EOS [0]{\spacefactor3000\relax}%
\providecommand \BibitemShut  [1]{\csname bibitem#1\endcsname}%
\let\auto@bib@innerbib\@empty
\bibitem [{\citenamefont {Accardi}\ \emph {et~al.}(2016)\citenamefont {Accardi}
  \emph {et~al.}}]{Accardi:2012qut}%
  \BibitemOpen
  \bibfield  {author} {\bibinfo {author} {\bibfnamefont {A.}~\bibnamefont
  {Accardi}} \emph {et~al.},\ }\href {\doibase 10.1140/epja/i2016-16268-9}
  {\bibfield  {journal} {\bibinfo  {journal} {The European Physical Journal A}\
  }\textbf {\bibinfo {volume} {52}},\ \bibinfo {pages} {268} (\bibinfo {year}
  {2016})}\BibitemShut {NoStop}%
\bibitem [{\citenamefont {Fazio}\ \emph {et~al.}(2012)\citenamefont {Fazio},
  \citenamefont {Fiore}, \citenamefont {Jenkovszky},\ and\ \citenamefont
  {Lavorini}}]{Fazio:2011ex}%
  \BibitemOpen
  \bibfield  {author} {\bibinfo {author} {\bibfnamefont {S.}~\bibnamefont
  {Fazio}}, \bibinfo {author} {\bibfnamefont {R.}~\bibnamefont {Fiore}},
  \bibinfo {author} {\bibfnamefont {L.}~\bibnamefont {Jenkovszky}}, \ and\
  \bibinfo {author} {\bibfnamefont {A.}~\bibnamefont {Lavorini}},\ }\href
  {\doibase 10.1103/PhysRevD.85.054009} {\bibfield  {journal} {\bibinfo
  {journal} {Phys. Rev.}\ }\textbf {\bibinfo {volume} {D85}},\ \bibinfo {pages}
  {054009} (\bibinfo {year} {2012})}\BibitemShut {NoStop}%
\bibitem [{\citenamefont {Martin}\ \emph
  {et~al.}(1997{\natexlab{a}})\citenamefont {Martin}, \citenamefont {Ryskin},\
  and\ \citenamefont {Teubner}}]{Martin:1996bp}%
  \BibitemOpen
  \bibfield  {author} {\bibinfo {author} {\bibfnamefont {A.~D.}\ \bibnamefont
  {Martin}}, \bibinfo {author} {\bibfnamefont {M.~G.}\ \bibnamefont {Ryskin}},
  \ and\ \bibinfo {author} {\bibfnamefont {T.}~\bibnamefont {Teubner}},\ }\href
  {\doibase 10.1103/PhysRevD.55.4329} {\bibfield  {journal} {\bibinfo
  {journal} {Phys. Rev.}\ }\textbf {\bibinfo {volume} {D55}},\ \bibinfo {pages}
  {4329} (\bibinfo {year} {1997}{\natexlab{a}})}\BibitemShut {NoStop}%
\bibitem [{\citenamefont {Golec-Biernat}\ and\ \citenamefont
  {Wusthoff}(1998)}]{GolecBiernat:1998js}%
  \BibitemOpen
  \bibfield  {author} {\bibinfo {author} {\bibfnamefont {K.~J.}\ \bibnamefont
  {Golec-Biernat}}\ and\ \bibinfo {author} {\bibfnamefont {M.}~\bibnamefont
  {Wusthoff}},\ }\href {\doibase 10.1103/PhysRevD.59.014017} {\bibfield
  {journal} {\bibinfo  {journal} {Phys. Rev.}\ }\textbf {\bibinfo {volume}
  {D59}},\ \bibinfo {pages} {014017} (\bibinfo {year} {1998})}\BibitemShut
  {NoStop}%
\bibitem [{\citenamefont {Kowalski}\ \emph {et~al.}(2006)\citenamefont
  {Kowalski}, \citenamefont {Motyka},\ and\ \citenamefont
  {Watt}}]{Kowalski:2006hc}%
  \BibitemOpen
  \bibfield  {author} {\bibinfo {author} {\bibfnamefont {H.}~\bibnamefont
  {Kowalski}}, \bibinfo {author} {\bibfnamefont {L.}~\bibnamefont {Motyka}}, \
  and\ \bibinfo {author} {\bibfnamefont {G.}~\bibnamefont {Watt}},\ }\href
  {\doibase 10.1103/PhysRevD.74.074016} {\bibfield  {journal} {\bibinfo
  {journal} {Phys. Rev.}\ }\textbf {\bibinfo {volume} {D74}},\ \bibinfo {pages}
  {074016} (\bibinfo {year} {2006})}\BibitemShut {NoStop}%
\bibitem [{\citenamefont {Jones}\ \emph {et~al.}(2013)\citenamefont {Jones},
  \citenamefont {Martin}, \citenamefont {Ryskin},\ and\ \citenamefont
  {Teubner}}]{Jones:2013pga}%
  \BibitemOpen
  \bibfield  {author} {\bibinfo {author} {\bibfnamefont {S.~P.}\ \bibnamefont
  {Jones}}, \bibinfo {author} {\bibfnamefont {A.~D.}\ \bibnamefont {Martin}},
  \bibinfo {author} {\bibfnamefont {M.~G.}\ \bibnamefont {Ryskin}}, \ and\
  \bibinfo {author} {\bibfnamefont {T.}~\bibnamefont {Teubner}},\ }\href
  {\doibase 10.1007/JHEP11(2013)085} {\bibfield  {journal} {\bibinfo  {journal}
  {JHEP}\ }\textbf {\bibinfo {volume} {2013}},\ \bibinfo {pages} {85} (\bibinfo
  {year} {2013})}\BibitemShut {NoStop}%
\bibitem [{\citenamefont {Jones}\ \emph {et~al.}(2016)\citenamefont {Jones},
  \citenamefont {Martin}, \citenamefont {Ryskin},\ and\ \citenamefont
  {Teubner}}]{Jones:2016ldq}%
  \BibitemOpen
  \bibfield  {author} {\bibinfo {author} {\bibfnamefont {S.~P.}\ \bibnamefont
  {Jones}}, \bibinfo {author} {\bibfnamefont {A.~D.}\ \bibnamefont {Martin}},
  \bibinfo {author} {\bibfnamefont {M.~G.}\ \bibnamefont {Ryskin}}, \ and\
  \bibinfo {author} {\bibfnamefont {T.}~\bibnamefont {Teubner}},\ }\href
  {\doibase 10.1140/epjc/s10052-016-4493-y} {\bibfield  {journal} {\bibinfo
  {journal} {The European Physical Journal C}\ }\textbf {\bibinfo {volume}
  {76}},\ \bibinfo {pages} {633} (\bibinfo {year} {2016})}\BibitemShut
  {NoStop}%
\bibitem [{\citenamefont {M{\"a}ntysaari}\ and\ \citenamefont
  {Venugopalan}(2017)}]{Mantysaari:2017slo}%
  \BibitemOpen
  \bibfield  {author} {\bibinfo {author} {\bibfnamefont {H.}~\bibnamefont
  {M{\"a}ntysaari}}\ and\ \bibinfo {author} {\bibfnamefont {R.}~\bibnamefont
  {Venugopalan}},\ }\href@noop {} {\  (\bibinfo {year} {2017})},\ \Eprint
  {http://arxiv.org/abs/1712.02508} {arXiv:1712.02508 [nucl-th]} \BibitemShut
  {NoStop}%
\bibitem [{\citenamefont {Adler}\ \emph {et~al.}(2002)\citenamefont {Adler}
  \emph {et~al.}}]{Adler:2002sc}%
  \BibitemOpen
  \bibfield  {author} {\bibinfo {author} {\bibfnamefont {C.}~\bibnamefont
  {Adler}} \emph {et~al.} (\bibinfo {collaboration} {STAR}),\ }\href {\doibase
  10.1103/PhysRevLett.89.272302} {\bibfield  {journal} {\bibinfo  {journal}
  {Phys. Rev. Lett.}\ }\textbf {\bibinfo {volume} {89}},\ \bibinfo {pages}
  {272302} (\bibinfo {year} {2002})}\BibitemShut {NoStop}%
\bibitem [{\citenamefont {Abelev}\ \emph {et~al.}(2008)\citenamefont {Abelev},
  \emph {et~al.}}]{Abelev:2007nb}%
  \BibitemOpen
  \bibfield  {author} {\bibinfo {author} {\bibfnamefont {B.~I.}\ \bibnamefont
  {Abelev}}, ,  \emph {et~al.} (\bibinfo {collaboration} {STAR
  Collaboration}),\ }\href {\doibase 10.1103/PhysRevC.77.034910} {\bibfield
  {journal} {\bibinfo  {journal} {Phys. Rev. C}\ }\textbf {\bibinfo {volume}
  {77}},\ \bibinfo {pages} {034910} (\bibinfo {year} {2008})}\BibitemShut
  {NoStop}%
\bibitem [{\citenamefont {Adamczyk}\ \emph {et~al.}(2017)\citenamefont
  {Adamczyk} \emph {et~al.}}]{Adamczyk:2017vfu}%
  \BibitemOpen
  \bibfield  {author} {\bibinfo {author} {\bibfnamefont {L.}~\bibnamefont
  {Adamczyk}} \emph {et~al.} (\bibinfo {collaboration} {STAR}),\ }\href
  {\doibase 10.1103/PhysRevC.96.054904} {\bibfield  {journal} {\bibinfo
  {journal} {Phys. Rev.}\ }\textbf {\bibinfo {volume} {C96}},\ \bibinfo {pages}
  {054904} (\bibinfo {year} {2017})}\BibitemShut {NoStop}%
\bibitem [{\citenamefont {Afanasiev}\ \emph {et~al.}(2009)\citenamefont
  {Afanasiev} \emph {et~al.}}]{Afanasiev:2009hy}%
  \BibitemOpen
  \bibfield  {author} {\bibinfo {author} {\bibfnamefont {S.}~\bibnamefont
  {Afanasiev}} \emph {et~al.} (\bibinfo {collaboration} {PHENIX}),\ }\href
  {\doibase 10.1016/j.physletb.2009.07.061} {\bibfield  {journal} {\bibinfo
  {journal} {Phys. Lett.}\ }\textbf {\bibinfo {volume} {B679}},\ \bibinfo
  {pages} {321} (\bibinfo {year} {2009})}\BibitemShut {NoStop}%
\bibitem [{\citenamefont {Abelev}\ \emph {et~al.}(2013)\citenamefont {Abelev}
  \emph {et~al.}}]{Abelev:2012ba}%
  \BibitemOpen
  \bibfield  {author} {\bibinfo {author} {\bibfnamefont {B.}~\bibnamefont
  {Abelev}} \emph {et~al.} (\bibinfo {collaboration} {ALICE}),\ }\href
  {\doibase 10.1016/j.physletb.2012.11.059} {\bibfield  {journal} {\bibinfo
  {journal} {Phys. Lett.}\ }\textbf {\bibinfo {volume} {B718}},\ \bibinfo
  {pages} {1273} (\bibinfo {year} {2013})}\BibitemShut {NoStop}%
\bibitem [{\citenamefont {Adam}\ \emph {et~al.}(2015)\citenamefont {Adam} \emph
  {et~al.}}]{Adam:2015gsa}%
  \BibitemOpen
  \bibfield  {author} {\bibinfo {author} {\bibfnamefont {J.}~\bibnamefont
  {Adam}} \emph {et~al.},\ }\href {\doibase 10.1007/JHEP09(2015)095} {\bibfield
   {journal} {\bibinfo  {journal} {JHEP}\ }\textbf {\bibinfo {volume} {2015}},\
  \bibinfo {pages} {95} (\bibinfo {year} {2015})}\BibitemShut {NoStop}%
\bibitem [{\citenamefont {Khachatryan}\ \emph {et~al.}(2017)\citenamefont
  {Khachatryan} \emph {et~al.}}]{Khachatryan:2016qhq}%
  \BibitemOpen
  \bibfield  {author} {\bibinfo {author} {\bibfnamefont {V.}~\bibnamefont
  {Khachatryan}} \emph {et~al.} (\bibinfo {collaboration} {CMS}),\ }\href
  {\doibase 10.1016/j.physletb.2017.07.001} {\bibfield  {journal} {\bibinfo
  {journal} {Phys. Lett.}\ }\textbf {\bibinfo {volume} {B772}},\ \bibinfo
  {pages} {489} (\bibinfo {year} {2017})}\BibitemShut {NoStop}%
\bibitem [{\citenamefont {Klein}(2017)}]{Klein:2017vua}%
  \BibitemOpen
  \bibfield  {author} {\bibinfo {author} {\bibfnamefont {S.~R.}\ \bibnamefont
  {Klein}},\ }\bibfield  {booktitle} {\emph {\bibinfo {booktitle} {{Proc. 26th
  Intl. Conf. on Ultra-relativistic Nucleus-Nucleus Collisions (Quark Matter
  2017)}}},\ }\href {\doibase 10.1016/j.nuclphysa.2017.05.098} {\bibfield
  {journal} {\bibinfo  {journal} {Nucl. Phys.}\ }\textbf {\bibinfo {volume}
  {A967}},\ \bibinfo {pages} {249} (\bibinfo {year} {2017})}\BibitemShut
  {NoStop}%
\bibitem [{\citenamefont {Toll}\ and\ \citenamefont
  {Ullrich}(2013)}]{Toll:2012mb}%
  \BibitemOpen
  \bibfield  {author} {\bibinfo {author} {\bibfnamefont {T.}~\bibnamefont
  {Toll}}\ and\ \bibinfo {author} {\bibfnamefont {T.}~\bibnamefont {Ullrich}},\
  }\href {\doibase 10.1103/PhysRevC.87.024913} {\bibfield  {journal} {\bibinfo
  {journal} {Phys. Rev. C}\ }\textbf {\bibinfo {volume} {87}},\ \bibinfo
  {pages} {024913} (\bibinfo {year} {2013})}\BibitemShut {NoStop}%
\bibitem [{\citenamefont {Kumano}\ and\ \citenamefont
  {Close}(1990)}]{Kumano:1989eh}%
  \BibitemOpen
  \bibfield  {author} {\bibinfo {author} {\bibfnamefont {S.}~\bibnamefont
  {Kumano}}\ and\ \bibinfo {author} {\bibfnamefont {F.~E.}\ \bibnamefont
  {Close}},\ }\href {\doibase 10.1103/PhysRevC.41.1855} {\bibfield  {journal}
  {\bibinfo  {journal} {Phys. Rev. C}\ }\textbf {\bibinfo {volume} {41}},\
  \bibinfo {pages} {1855} (\bibinfo {year} {1990})}\BibitemShut {NoStop}%
\bibitem [{\citenamefont {Kitagaki}\ \emph {et~al.}(1988)\citenamefont
  {Kitagaki} \emph {et~al.}}]{Kitagaki:1988wc}%
  \BibitemOpen
  \bibfield  {author} {\bibinfo {author} {\bibfnamefont {T.}~\bibnamefont
  {Kitagaki}} \emph {et~al.},\ }\href {\doibase
  https://doi.org/10.1016/0370-2693(88)91483-9} {\bibfield  {journal} {\bibinfo
   {journal} {Physics Letters B}\ }\textbf {\bibinfo {volume} {214}},\ \bibinfo
  {pages} {281 } (\bibinfo {year} {1988})}\BibitemShut {NoStop}%
\bibitem [{\citenamefont {Emel'yanov}\ \emph {et~al.}(1998)\citenamefont
  {Emel'yanov}, \citenamefont {Khodinov}, \citenamefont {Klein},\ and\
  \citenamefont {Vogt}}]{Emelyanov:1998yul}%
  \BibitemOpen
  \bibfield  {author} {\bibinfo {author} {\bibfnamefont {V.}~\bibnamefont
  {Emel'yanov}}, \bibinfo {author} {\bibfnamefont {A.}~\bibnamefont
  {Khodinov}}, \bibinfo {author} {\bibfnamefont {S.~R.}\ \bibnamefont {Klein}},
  \ and\ \bibinfo {author} {\bibfnamefont {R.}~\bibnamefont {Vogt}},\ }\href
  {\doibase 10.1103/PhysRevLett.81.1801} {\bibfield  {journal} {\bibinfo
  {journal} {Phys. Rev. Lett.}\ }\textbf {\bibinfo {volume} {81}},\ \bibinfo
  {pages} {1801} (\bibinfo {year} {1998})}\BibitemShut {NoStop}%
\bibitem [{\citenamefont {Montag}(2017)}]{eRHIC}%
  \BibitemOpen
  \bibfield  {author} {\bibinfo {author} {\bibfnamefont {C.}~\bibnamefont
  {Montag}},\ }\href {https://agenda.infn.it/contributionDisplay.py?
  sessionId=14\&contribId=83\&confId=13037} {}\bibinfo {howpublished}
  {presented at the EIC User Group Meeting 2017, Trieste, Italy} (\bibinfo
  {year} {2017})\BibitemShut {NoStop}%
\bibitem [{\citenamefont {Morozov}(2017)}]{JLEIC}%
  \BibitemOpen
  \bibfield  {author} {\bibinfo {author} {\bibfnamefont {V.}~\bibnamefont
  {Morozov}},\ }\href {https://agenda.infn.it/contributionDisplay.py?
  sessionId=14\&contribId=82\&confId=13037} {}\bibinfo {howpublished}
  {presented at the ``EIC User Group Meeting 2017 Trieste, Italy} (\bibinfo
  {year} {2017})\BibitemShut {NoStop}%
\bibitem [{\citenamefont {Fernandez}\ \emph {et~al.}(2012)\citenamefont
  {Fernandez} \emph {et~al.}}]{AbelleiraFernandez:2012cc}%
  \BibitemOpen
  \bibfield  {author} {\bibinfo {author} {\bibfnamefont {J.~L.~A.}\
  \bibnamefont {Fernandez}} \emph {et~al.} (\bibinfo {collaboration} {LHeC
  Study Group}),\ }\href {http://stacks.iop.org/0954-3899/39/i=7/a=075001}
  {\bibfield  {journal} {\bibinfo  {journal} {JPhysG}\ }\textbf {\bibinfo
  {volume} {39}},\ \bibinfo {pages} {075001} (\bibinfo {year}
  {2012})}\BibitemShut {NoStop}%
\bibitem [{\citenamefont {Klein}\ \emph {et~al.}(2017)\citenamefont {Klein},
  \citenamefont {Nystrand}, \citenamefont {Seger}, \citenamefont {Gorbunov},\
  and\ \citenamefont {Butterworth}}]{Klein:2016yzr}%
  \BibitemOpen
  \bibfield  {author} {\bibinfo {author} {\bibfnamefont {S.~R.}\ \bibnamefont
  {Klein}}, \bibinfo {author} {\bibfnamefont {J.}~\bibnamefont {Nystrand}},
  \bibinfo {author} {\bibfnamefont {J.}~\bibnamefont {Seger}}, \bibinfo
  {author} {\bibfnamefont {Y.}~\bibnamefont {Gorbunov}}, \ and\ \bibinfo
  {author} {\bibfnamefont {J.}~\bibnamefont {Butterworth}},\ }\href {\doibase
  https://doi.org/10.1016/j.cpc.2016.10.016} {\bibfield  {journal} {\bibinfo
  {journal} {Computer Physics Communications}\ }\textbf {\bibinfo {volume}
  {212}},\ \bibinfo {pages} {258 } (\bibinfo {year} {2017})}\BibitemShut
  {NoStop}%
\bibitem [{\citenamefont {Abelev}\ \emph {et~al.}(2014)\citenamefont {Abelev}
  \emph {et~al.}}]{PhysRevLett.113.232504}%
  \BibitemOpen
  \bibfield  {author} {\bibinfo {author} {\bibfnamefont {B.}~\bibnamefont
  {Abelev}} \emph {et~al.} (\bibinfo {collaboration} {ALICE Collaboration}),\
  }\href {\doibase 10.1103/PhysRevLett.113.232504} {\bibfield  {journal}
  {\bibinfo  {journal} {Phys. Rev. Lett.}\ }\textbf {\bibinfo {volume} {113}},\
  \bibinfo {pages} {232504} (\bibinfo {year} {2014})}\BibitemShut {NoStop}%
\bibitem [{\citenamefont {Toll}\ and\ \citenamefont
  {Ullrich}(2014)}]{Toll:2013gda}%
  \BibitemOpen
  \bibfield  {author} {\bibinfo {author} {\bibfnamefont {T.}~\bibnamefont
  {Toll}}\ and\ \bibinfo {author} {\bibfnamefont {T.}~\bibnamefont {Ullrich}},\
  }\href {\doibase https://doi.org/10.1016/j.cpc.2014.03.010} {\bibfield
  {journal} {\bibinfo  {journal} {Computer Physics Communications}\ }\textbf
  {\bibinfo {volume} {185}},\ \bibinfo {pages} {1835 } (\bibinfo {year}
  {2014})}\BibitemShut {NoStop}%
\bibitem [{\citenamefont {Crittenden}(1997)}]{Crittenden:1997yz}%
  \BibitemOpen
  \bibfield  {author} {\bibinfo {author} {\bibfnamefont {J.~A.}\ \bibnamefont
  {Crittenden}},\ }\href@noop {} {\  (\bibinfo {year} {1997})},\ \Eprint
  {http://arxiv.org/abs/hep-ex/9704009} {arXiv:hep-ex/9704009 [hep-ex]}
  \BibitemShut {NoStop}%
\bibitem [{\citenamefont {Budnev}\ \emph {et~al.}(1975)\citenamefont {Budnev},
  \citenamefont {Ginzburg}, \citenamefont {Meledin},\ and\ \citenamefont
  {Serbo}}]{Budnev1975181}%
  \BibitemOpen
  \bibfield  {author} {\bibinfo {author} {\bibfnamefont {V.}~\bibnamefont
  {Budnev}}, \bibinfo {author} {\bibfnamefont {I.}~\bibnamefont {Ginzburg}},
  \bibinfo {author} {\bibfnamefont {G.}~\bibnamefont {Meledin}}, \ and\
  \bibinfo {author} {\bibfnamefont {V.}~\bibnamefont {Serbo}},\ }\href
  {\doibase https://doi.org/10.1016/0370-1573(75)90009-5} {\bibfield  {journal}
  {\bibinfo  {journal} {Phys. Rep.}\ }\textbf {\bibinfo {volume} {15}},\
  \bibinfo {pages} {181 } (\bibinfo {year} {1975})}\BibitemShut {NoStop}%
\bibitem [{\citenamefont {Kopeliovich}\ \emph {et~al.}(2002)\citenamefont
  {Kopeliovich}, \citenamefont {Nemchik}, \citenamefont {Sch\"afer},\ and\
  \citenamefont {Tarasov}}]{PhysRevC.65.035201}%
  \BibitemOpen
  \bibfield  {author} {\bibinfo {author} {\bibfnamefont {B.~Z.}\ \bibnamefont
  {Kopeliovich}}, \bibinfo {author} {\bibfnamefont {J.}~\bibnamefont
  {Nemchik}}, \bibinfo {author} {\bibfnamefont {A.}~\bibnamefont {Sch\"afer}},
  \ and\ \bibinfo {author} {\bibfnamefont {A.~V.}\ \bibnamefont {Tarasov}},\
  }\href {\doibase 10.1103/PhysRevC.65.035201} {\bibfield  {journal} {\bibinfo
  {journal} {Phys. Rev. C}\ }\textbf {\bibinfo {volume} {65}},\ \bibinfo
  {pages} {035201} (\bibinfo {year} {2002})}\BibitemShut {NoStop}%
\bibitem [{\citenamefont {Klein}\ and\ \citenamefont
  {Nystrand}(1999)}]{Klein:1999qj}%
  \BibitemOpen
  \bibfield  {author} {\bibinfo {author} {\bibfnamefont {S.~R.}\ \bibnamefont
  {Klein}}\ and\ \bibinfo {author} {\bibfnamefont {J.}~\bibnamefont
  {Nystrand}},\ }\href {\doibase 10.1103/PhysRevC.60.014903} {\bibfield
  {journal} {\bibinfo  {journal} {Phys. Rev. C}\ }\textbf {\bibinfo {volume}
  {60}},\ \bibinfo {pages} {014903} (\bibinfo {year} {1999})}\BibitemShut
  {NoStop}%
\bibitem [{\citenamefont {Adloff}\ \emph {et~al.}(2000)\citenamefont {Adloff}
  \emph {et~al.}}]{Adloffetal.2000}%
  \BibitemOpen
  \bibfield  {author} {\bibinfo {author} {\bibfnamefont {C.}~\bibnamefont
  {Adloff}} \emph {et~al.} (\bibinfo {collaboration} {H1}),\ }\href {\doibase
  10.1007/s100520000150} {\bibfield  {journal} {\bibinfo  {journal} {The
  European Physical Journal C}\ }\textbf {\bibinfo {volume} {13}},\ \bibinfo
  {pages} {371} (\bibinfo {year} {2000})}\BibitemShut {NoStop}%
\bibitem [{\citenamefont {Aaron}\ \emph {et~al.}(2010)\citenamefont {Aaron}
  \emph {et~al.}}]{Aaron2010}%
  \BibitemOpen
  \bibfield  {author} {\bibinfo {author} {\bibfnamefont {F.~D.}\ \bibnamefont
  {Aaron}} \emph {et~al.},\ }\href {\doibase 10.1007/JHEP05(2010)032}
  {\bibfield  {journal} {\bibinfo  {journal} {JHEP}\ }\textbf {\bibinfo
  {volume} {2010}},\ \bibinfo {pages} {32} (\bibinfo {year}
  {2010})}\BibitemShut {NoStop}%
\bibitem [{\citenamefont {Nemchik}\ \emph {et~al.}(1997)\citenamefont
  {Nemchik}, \citenamefont {Nikolaev}, \citenamefont {Predazzi},\ and\
  \citenamefont {Zakharov}}]{Nemchik:1996cw}%
  \BibitemOpen
  \bibfield  {author} {\bibinfo {author} {\bibfnamefont {J.}~\bibnamefont
  {Nemchik}}, \bibinfo {author} {\bibfnamefont {N.~N.}\ \bibnamefont
  {Nikolaev}}, \bibinfo {author} {\bibfnamefont {E.}~\bibnamefont {Predazzi}},
  \ and\ \bibinfo {author} {\bibfnamefont {B.~G.}\ \bibnamefont {Zakharov}},\
  }\href {\doibase 10.1007/s002880050448} {\bibfield  {journal} {\bibinfo
  {journal} {Z. Phys.}\ }\textbf {\bibinfo {volume} {C75}},\ \bibinfo {pages}
  {71} (\bibinfo {year} {1997})}\BibitemShut {NoStop}%
\bibitem [{\citenamefont {Martin}\ \emph
  {et~al.}(1997{\natexlab{b}})\citenamefont {Martin}, \citenamefont {Ryskin},\
  and\ \citenamefont {Teubner}}]{Martin:1997sh}%
  \BibitemOpen
  \bibfield  {author} {\bibinfo {author} {\bibfnamefont {A.~D.}\ \bibnamefont
  {Martin}}, \bibinfo {author} {\bibfnamefont {M.~G.}\ \bibnamefont {Ryskin}},
  \ and\ \bibinfo {author} {\bibfnamefont {T.}~\bibnamefont {Teubner}},\ }\href
  {\doibase 10.1103/PhysRevD.56.3007} {\bibfield  {journal} {\bibinfo
  {journal} {Phys. Rev.}\ }\textbf {\bibinfo {volume} {D56}},\ \bibinfo {pages}
  {3007} (\bibinfo {year} {1997}{\natexlab{b}})}\BibitemShut {NoStop}%
\bibitem [{\citenamefont {Breitweg}\ \emph {et~al.}(1999)\citenamefont
  {Breitweg} \emph {et~al.}}]{Breitwegetal.1999}%
  \BibitemOpen
  \bibfield  {author} {\bibinfo {author} {\bibfnamefont {J.}~\bibnamefont
  {Breitweg}} \emph {et~al.} (\bibinfo {collaboration} {ZEUS}),\ }\href
  {\doibase 10.1007/s100529901051} {\bibfield  {journal} {\bibinfo  {journal}
  {The European Physical Journal C}\ }\textbf {\bibinfo {volume} {6}},\
  \bibinfo {pages} {603} (\bibinfo {year} {1999})}\BibitemShut {NoStop}%
\bibitem [{\citenamefont {Frankfurt}\ \emph {et~al.}(2002)\citenamefont
  {Frankfurt}, \citenamefont {Strikman},\ and\ \citenamefont
  {Zhalov}}]{Frankfurt:2002wc}%
  \BibitemOpen
  \bibfield  {author} {\bibinfo {author} {\bibfnamefont {L.}~\bibnamefont
  {Frankfurt}}, \bibinfo {author} {\bibfnamefont {M.}~\bibnamefont {Strikman}},
  \ and\ \bibinfo {author} {\bibfnamefont {M.}~\bibnamefont {Zhalov}},\ }\href
  {\doibase https://doi.org/10.1016/S0370-2693(02)01882-8} {\bibfield
  {journal} {\bibinfo  {journal} {Physics Letters B}\ }\textbf {\bibinfo
  {volume} {537}},\ \bibinfo {pages} {51 } (\bibinfo {year}
  {2002})}\BibitemShut {NoStop}%
\bibitem [{\citenamefont {Frankfurt}\ \emph {et~al.}(2016)\citenamefont
  {Frankfurt}, \citenamefont {Guzey}, \citenamefont {Strikman},\ and\
  \citenamefont {Zhalov}}]{Frankfurt:2015cwa}%
  \BibitemOpen
  \bibfield  {author} {\bibinfo {author} {\bibfnamefont {L.}~\bibnamefont
  {Frankfurt}}, \bibinfo {author} {\bibfnamefont {V.}~\bibnamefont {Guzey}},
  \bibinfo {author} {\bibfnamefont {M.}~\bibnamefont {Strikman}}, \ and\
  \bibinfo {author} {\bibfnamefont {M.}~\bibnamefont {Zhalov}},\ }\href
  {\doibase 10.1016/j.physletb.2015.11.012} {\bibfield  {journal} {\bibinfo
  {journal} {Phys. Lett.}\ }\textbf {\bibinfo {volume} {B752}},\ \bibinfo
  {pages} {51} (\bibinfo {year} {2016})}\BibitemShut {NoStop}%
\bibitem [{\citenamefont {Klein}\ and\ \citenamefont
  {Nystrand}(2000)}]{Klein:1999gv}%
  \BibitemOpen
  \bibfield  {author} {\bibinfo {author} {\bibfnamefont {S.~R.}\ \bibnamefont
  {Klein}}\ and\ \bibinfo {author} {\bibfnamefont {J.}~\bibnamefont
  {Nystrand}},\ }\href {\doibase 10.1103/PhysRevLett.84.2330} {\bibfield
  {journal} {\bibinfo  {journal} {Phys. Rev. Lett.}\ }\textbf {\bibinfo
  {volume} {84}},\ \bibinfo {pages} {2330} (\bibinfo {year}
  {2000})}\BibitemShut {NoStop}%
\bibitem [{\citenamefont {Drees}\ \emph {et~al.}(1989)\citenamefont {Drees},
  \citenamefont {Ellis},\ and\ \citenamefont {Zeppenfeld}}]{Drees:1989vq}%
  \BibitemOpen
  \bibfield  {author} {\bibinfo {author} {\bibfnamefont {M.}~\bibnamefont
  {Drees}}, \bibinfo {author} {\bibfnamefont {J.~R.}\ \bibnamefont {Ellis}}, \
  and\ \bibinfo {author} {\bibfnamefont {D.}~\bibnamefont {Zeppenfeld}},\
  }\href {\doibase 10.1016/0370-2693(89)91632-8} {\bibfield  {journal}
  {\bibinfo  {journal} {Phys. Lett.}\ }\textbf {\bibinfo {volume} {B223}},\
  \bibinfo {pages} {454} (\bibinfo {year} {1989})}\BibitemShut {NoStop}%
\bibitem [{\citenamefont {Armesto}\ and\ \citenamefont
  {Rezaeian}(2014)}]{Armesto:2014sma}%
  \BibitemOpen
  \bibfield  {author} {\bibinfo {author} {\bibfnamefont {N.}~\bibnamefont
  {Armesto}}\ and\ \bibinfo {author} {\bibfnamefont {A.~H.}\ \bibnamefont
  {Rezaeian}},\ }\href {\doibase 10.1103/PhysRevD.90.054003} {\bibfield
  {journal} {\bibinfo  {journal} {Phys. Rev.}\ }\textbf {\bibinfo {volume}
  {D90}},\ \bibinfo {pages} {054003} (\bibinfo {year} {2014})}\BibitemShut
  {NoStop}%
\bibitem [{\citenamefont {Schildknecht}\ \emph {et~al.}(1999)\citenamefont
  {Schildknecht}, \citenamefont {Schuler},\ and\ \citenamefont
  {Surrow}}]{SCHILDKNECHT1999328}%
  \BibitemOpen
  \bibfield  {author} {\bibinfo {author} {\bibfnamefont {D.}~\bibnamefont
  {Schildknecht}}, \bibinfo {author} {\bibfnamefont {G.~A.}\ \bibnamefont
  {Schuler}}, \ and\ \bibinfo {author} {\bibfnamefont {B.}~\bibnamefont
  {Surrow}},\ }\href {\doibase https://doi.org/10.1016/S0370-2693(99)00052-0}
  {\bibfield  {journal} {\bibinfo  {journal} {Physics Letters B}\ }\textbf
  {\bibinfo {volume} {449}},\ \bibinfo {pages} {328 } (\bibinfo {year}
  {1999})}\BibitemShut {NoStop}%
\bibitem [{\citenamefont {Chekanov}\ \emph {et~al.}(2004)\citenamefont
  {Chekanov} \emph {et~al.}}]{Chekanov20043}%
  \BibitemOpen
  \bibfield  {author} {\bibinfo {author} {\bibfnamefont {S.}~\bibnamefont
  {Chekanov}} \emph {et~al.},\ }\href {\doibase
  https://doi.org/10.1016/j.nuclphysb.2004.06.034} {\bibfield  {journal}
  {\bibinfo  {journal} {Nuclear Physics B}\ }\textbf {\bibinfo {volume}
  {695}},\ \bibinfo {pages} {3 } (\bibinfo {year} {2004})}\BibitemShut
  {NoStop}%
\bibitem [{\citenamefont {Guzey}\ \emph {et~al.}(2017)\citenamefont {Guzey},
  \citenamefont {Strikman},\ and\ \citenamefont {Zhalov}}]{Guzey:2016qwo}%
  \BibitemOpen
  \bibfield  {author} {\bibinfo {author} {\bibfnamefont {V.}~\bibnamefont
  {Guzey}}, \bibinfo {author} {\bibfnamefont {M.}~\bibnamefont {Strikman}}, \
  and\ \bibinfo {author} {\bibfnamefont {M.}~\bibnamefont {Zhalov}},\ }\href
  {\doibase 10.1103/PhysRevC.95.025204} {\bibfield  {journal} {\bibinfo
  {journal} {Phys. Rev.}\ }\textbf {\bibinfo {volume} {C95}},\ \bibinfo {pages}
  {025204} (\bibinfo {year} {2017})}\BibitemShut {NoStop}%
\bibitem [{\citenamefont {Bultmann}\ \emph {et~al.}(2004)\citenamefont
  {Bultmann} \emph {et~al.}}]{Bultmann:2004ke}%
  \BibitemOpen
  \bibfield  {author} {\bibinfo {author} {\bibfnamefont {S.}~\bibnamefont
  {Bultmann}} \emph {et~al.},\ }\bibfield  {booktitle} {\emph {\bibinfo
  {booktitle} {{Instrumentation. Proc., 10th Intl. Conf. VCI 10, Vienna,
  Austria, February 16-21, 2004}}},\ }\href {\doibase
  10.1016/j.nima.2004.07.162} {\bibfield  {journal} {\bibinfo  {journal} {Nucl.
  Instrum. Meth.}\ }\textbf {\bibinfo {volume} {A535}},\ \bibinfo {pages} {415}
  (\bibinfo {year} {2004})}\BibitemShut {NoStop}%
\bibitem [{\citenamefont {Klein}(2014)}]{Klein:2014xoa}%
  \BibitemOpen
  \bibfield  {author} {\bibinfo {author} {\bibfnamefont {S.~R.}\ \bibnamefont
  {Klein}},\ }\href {\doibase 10.1103/PhysRevSTAB.17.121003} {\bibfield
  {journal} {\bibinfo  {journal} {Phys. Rev. ST Accel. Beams}\ }\textbf
  {\bibinfo {volume} {17}},\ \bibinfo {pages} {121003} (\bibinfo {year}
  {2014})}\BibitemShut {NoStop}%
\end{thebibliography}%

\end{document}